\documentclass[a4paper,titlepage,12pt]{article}
\usepackage[top=2.5cm, bottom=2.5cm, left=2.5cm, right=2.5cm]{geometry}
\usepackage[dvips]{graphicx}
\usepackage{subfigure}
\usepackage[cmex10]{amsmath}
\usepackage{algorithmic}
\usepackage{mdwmath}
\usepackage{mdwtab}

\begin{document}
\title{Information-Theoretic Study on Routing Path Selection in Two-Way Relay Networks}

\author{Shanshan Wu, Xudong Wang}
\maketitle 
\begin{abstract}
Two-way relaying is a promising technique to improve network throughput. However, how to apply it to a wireless network remains an unresolved issue. Particularly, challenges lie in the joint design between the physical layer and the routing protocol. Applying an existing routing protocol to a two-way relay network can easily compromise the advantages of two-way relaying. Considering routing path selection and two-way relaying together can be formulated as a network optimization problem, but it is usually NP-hard. In this paper, we take a different approach to study routing path selection for two-way relay networks. Instead of solving the joint optimization problem, we study the fundamental characteristics of a routing path consisting of multihop two-way relaying nodes. Information theoretical analysis is carried out to derive bandwidth efficiency and energy efficiency of a routing path in a two-way relay network. Such analysis provides a framework of routing path selection by considering bandwidth efficiency, energy efficiency and latency subject to physical layer constraints such as the transmission rate, transmission power, path loss exponent, path length and the number of relays. This framework provides insightful guidelines on routing protocol design of a two-way relay network. Our analytical framework and insights are illustrated by extensive numerical results.
\end{abstract}

\section{Introduction}
\begin{figure}[ht]
\centering
\includegraphics[scale=0.6]{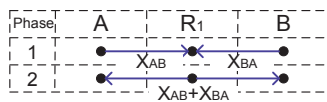}
\caption{Illustration of a TWRC.}\label{TWRC}
\end{figure}

\indent
Two-way relay channel (TWRC) improve throughput by exploiting bi-directional interference~\cite{proceedings}. A typical model for TWRC contains three nodes, as shown in Fig.~\ref{TWRC}, where A and B want to exchange data via relay R$_1$, assuming that all nodes operate in a half-duplex mode. Taking amplify-and-forward (AF) TWRC as an example, A and B transmit their packets simultaneously to relay in the first time slot. Then R$_1$ amplifies and broadcasts the superimposed waveforms in the second time slot. After receiving it, A and B subtract its own signal to obtain their desired data. By utilizing interference instead of regarding it as noise, TWRC enables A and B to exchange data in two time slots, which is only half of the time needed in the conventional routing scheme. Therefore, two-way relaying has become a promising technique to improve network performance and has been extensively studied recently, e.g., ~\cite{proceedings} -~\cite{AF-ANC-Even}. Although there are different two-way relay techniques~\cite{proceedings}~\cite{compute}~\cite{anti}, this paper is focused on AF TWRC for the following two reasons. Firstly, AF is simple to implement and insensitive to the environment change, such as variations in the coding schemes or the modulation methods~\cite{half}~\cite{denoising_map}. Secondly, the performance of AF is comparable with other techniques. In~\cite{proceedings}, we see that at low transmission power, it has acceptable performance; at high transmission power, its performance is even better than most techniques.\\
\begin{figure}
\centering
\includegraphics[scale=0.6]{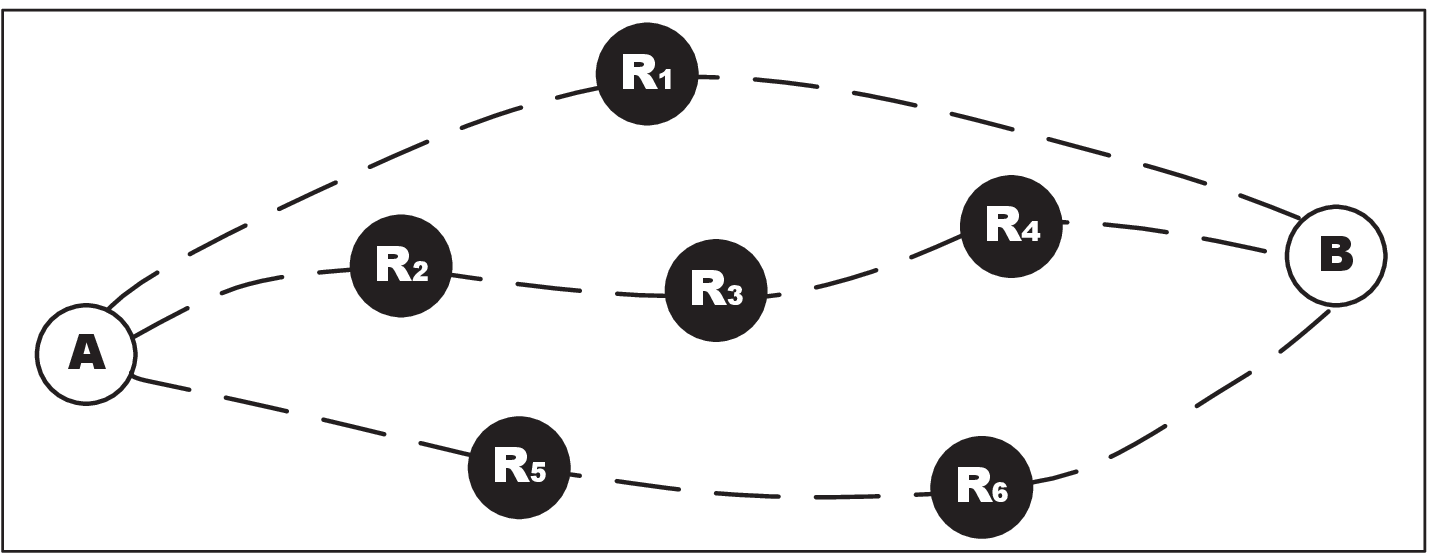}
\caption{Illustration of a wireless network: A and B want to exchange data, but there are more than one relays in between. How to select an end-to-end routing path so that the route offers best performance in the presence of AF TWRCs?}\label{network}
\end{figure}
\indent
In spite of the advantages of TWRC, how to efficiently use this technique to acquire performance gains in a wireless network remains a challenging problem. For example, in Fig.~\ref{network}, where A and B want to exchange data and there are more than one relays in between, if we want to transmit through AF TWRCs, one crucial question is that how to select an end-to-end routing path in the presence of TWRCs. To the best of our knowledge, no research work is reported on end-to-end routing in an AF two-way relay network. Thus, we are motivated to tackle this problem. In this paper, we develop a framework to analyze and compare the performance of different routes, from a perspective of information theory. Therefore, the three most fundamental parameters, i.e., energy efficiency (EE), bandwidth efficiency (BE), and latency, are considered as the metrics for evaluating the performance of a routing path. Conventional routing metrics, such as hop count, ETX~\cite{EXT}, ETT~\cite{ETT}, are not applicable to our framework. EE and BE are measures of how efficiently the network utilizes energy (including both transmission and processing energy) and bandwidth to transmit data. In this paper, we use a model for processing energy instead of assuming it as a constant as in~\cite{EE-BE} -~\cite{ismore}. Therefore, we are able to see with more accuracy how processing energy influences the network performance. Note that BE also characterizes the end-to-end rate in bit per channel use. Latency denotes the time elapse when data traverse the network. In other words, it measures the delay experienced by each bit before it reaches the destination. High EE, high BE, and low latency are the desired properties for a transmission scheme. However, normally they cannot be achieved simultaneously since tradeoff exists between them~\cite{verdu}~\cite{neely}: we need to sacrifice one of them in order to gain the improvement in the other.\\
\indent
The ultimate goal of this paper is to provide insights towards routing protocol design considering the fundamental requirements of BE, EE and latency. To build a framework for studying the performance of different routing paths from the perspective of BE and EE, we adopt a bottom-up method, beginning with analysis of a traditional three-node AF TWRC. Interestingly, given BE, we find a threshold for path loss exponent such that, when it is above the threshold, TWRC with relay located in the middle consumes the smallest energy. We also derive the power allocation associated with that condition of smallest energy consumption. After the simple case is studied, we extend the approach of analysis to an AF two-way relay network, where multiple pairs of nodes want to exchange data through more than one relays. With perfect scheduling, this two-way relay network is decomposed into a number of small networks each with only one source-destination (SD) pair. Thus we only need to consider one of those small networks, with an aim at selecting an optimal routing path for that particular SD pair. To better use the results gained in the previous analysis of three-node AF TWRC, we assume that each routing path has a small curvature and equi-spaced relays. Under this assumption, we formulate the three metrics, i.e., BE, EE, and latency, for routes with different number of relays. Numerical analysis is then carried out to analyze and compare the performance of different routes. Specifically, we find that the output parameters BE and EE of a route of a given SD pair are determined by the path length, number of relays and processing energy model. Since high EE, high BE, and low latency cannot be achieved simultaneously~\cite{verdu}~\cite{neely}, we define a general objective function integrating those three metrics. The optimal route can be found through this function.\\
The following contributions are made in this paper:
\begin{itemize}
\item We build an information theoretic framework to select an optimal routing path in an AF two-way relay network that provides the best tradeoff between EE, BE, and latency.
\item For AF TWRC, We find a threshold as a function of the transmission rate. When the path loss exponent surpasses the threshold, locating relay in the middle leads to the lowest energy consumption.
\item A power allocation scheme is developed to different routes. It allows each route to consume the smallest energy while still achieve the same transmission rate.
\end{itemize}

This paper is organized as follows. Section II formulate the performance metrics for a traditional three-node AF TWRC, and analyzes how relay's location impacts the total energy consumption. Section III presents the network model and compares two schemes that enable multi-hop transmission through TWRCs. In Section IV we extend the analysis frame to an AF two-way relay network, based on the Hop-by-Hop scheme. Section V studies the performance of different routes based on numerical results, and provides insightful observations towards routing protocol design. This paper concludes in Section VI.\\
\indent
The discussions in this paper are based on the following notations.
\begin{itemize}
\item $x_i$: symbol transmitted by node i
\item $y_i$: symbol received by node i
\item $z_i$: additive white noise at node i, $\mathcal{CN}(0,N_0)$
\item $\alpha$: path loss exponent
\item $\beta$: denote how many times relay amplifies the received signal
\item $R$: transmission rate at each link (bit/channel use)
\item $R_i$: the ith relay
\item $\gamma$: bandwidth efficiency (bit/s/Hz)
\item $R_e$: end-to-end rate (bit/channel use)
\item $d_{route}$: length of a route (m)
\item $P_i$: transmission energy per channel use at node i (J/channel use)
\item $P_{proc,k}$: average processing energy per channel use for a route with k relays (J/channel use)
\item $\xi$: energy efficiency (bit/J)
\item $k$: total number of relays along a route, $k=0,1,...,6$
\end{itemize}

\section{Three-Node Amplify-and-Forward TWRC}
\indent
In this section, we focus on a simple route with only one relay, i.e., a traditional three-node AF TWRC. We first introduce a model for computing processing energy, based on which the system's performance measures could be derived. We are then interested in finding when the route gives best performance, by varying relay's location and the amplification factor. Previous research has found that AF TWRC with relay in the middle gives the highest sum rate~\cite{full} and lowest outage probability~\cite{outage}, with a given power allocation and fixed path loss exponent. But in this section we let both of the power allocation and path loss exponent be variables, and find that relay in the middle does not always provide the minimum energy consumption. One related work, i.e.,~\cite{provision}, considers the power provision and relay placement problem for AF TWRC under Rayleigh fading and given outage probability, reaching a conclusion that relay is best positioned at the middle point to achieve lowest energy consumption. But our analysis is performed under different assumptions, i.e., large scale fading and capacity-achieving coding, which consequently leads to a different conclusion: whether relay in the middle is the most energy efficient depends on the relation between path loss exponent and transmission rate.
\subsection{Model for Processing Energy}
\indent
When data is transmitted from source to destination, energy is dissipated at two places: in the radio hardware (i.e., processing energy, consumed at both source and destination) and in the EM waves (i.e., transmission energy, only consumed at the source node). For the processing energy, we assume a similar model as in~\cite{processing}, which is simpler than that in~\cite{processing}, yet still captures the main characteristics of the circuit power consumption. The total processing energy $P_{proc}$ for a transmitter-receiver pair is
\begin{equation}
\begin{split}
&\text{Transmitter:}\; P_{T}=(\frac{1}{\eta}-1)P_{tr}+P_{T0}, \\
&\text{Receiver:}\; P_{R}= P_{R0},\end{split} \label{total_energy}
\end{equation}
where $P_T$ and $P_R$ denote processing energy dissipated at the transmitter and receiver; $P_{tr}$ is the transmission energy carried by the EM waves; $\eta$ is a constant representing the drain efficiency of the power amplifier (PA); $P_{T0}$ and $P_{R0}$ are constants representing the energy consumption in radio electronics except PA, such as ADC, DAC, LNA, etc.
\subsection{Performance Measures}
\indent
For the derivation in this section and the following section, let $x_i$ and $y_i$ denote the transmitted and received symbols by node $i$, let $z_i$ be the noise at node $i$ with a distribution of $\mathcal{CN}(0,N_0)$, let $P_i$ be the transmission energy per channel use at node $i$. Note that we do not distinguish between symbols transmitted or received in different time slots, for no ambiguity will occur during the derivation. Assuming that each node operates in half-duplex mode. As shown in Fig.~\ref{TWRC}, A and B will transmit their packets simultaneously to relay in the first time slot,, i.e.,
\begin{equation}
y_R=h_{A}x_A+h_{B}x_B+z_R,
\end{equation}
where $h_A$ and $h_B$ are the channel gains of the channels between A and relay, B and relay, respectively (assuming symmetric channels). After receiving the overlapped waveform, relay amplifies and broadcasts it in the next time slot, i.e.,
\begin{equation}
x_R=\beta(h_{A}x_A+h_{B}x_B+z_R),
\end{equation}
where $\beta$ denotes how many times relay amplifies its received signal. After receiving it, A and B will perform the so called "self-cancellation" to extract their desired data, i.e.,
\begin{equation}
y_A-h_A\beta h_Ax_A=h_A\beta h_Bx_B+h_A\beta z_R+z_A,
\end{equation}
\begin{equation}
y_B-h_B\beta h_Bx_B=h_B\beta h_Ax_A+h_B\beta z_R+z_B.
\end{equation}
Assuming that transmission is done by capacity-achieving codes and a common transmission rate at each link. Then
\begin{equation}
\begin{split}R&=\log_2(1+\frac{|h_A|^2|\beta|^2|h_B|^2P_B}{(|h_A|^2|\beta|^2+1)N_0})\\
&=\log_2(1+\frac{|h_B|^2|\beta|^2|h_A|^2P_A}{(|h_B|^2|\beta|^2+1)N_0})\end{split}, \label{rate}
\end{equation}
and
\begin{equation}
|\beta|=\sqrt{\frac{P_R}{|h_A|^2P_A+|h_B|^2P_B+N_0}}.\label{beta}
\end{equation}
\subsubsection{Latency}
A and B exchanges data every two time slots, so the latency experienced by each bit is $2$ time slots/bit.
\subsubsection{Bandwidth Efficiency (BE)}
Assuming that A and B exchange data with a common transmission rate $R$ (bit/channel use) at each link, the end-to-end rate $R_e$ (bit/channel use) of the three-node system is $R_e=R$. Given that each complex channel dimension occupies $\rho$ seconds $\times$ hertz, where $\rho$ is a constant depending on the techniques in the physical layer, then the BE (bit/s/Hz) of this system is $\gamma=R_e/\rho$~\cite{verdu}. In this paper, we assume $\rho=1$, which corresponds to the maximum BE
\begin{equation}
\gamma=R_e/\rho=R_e =R. \label{BE1}
\end{equation}
\subsubsection{Energy Efficiency (EE)}
Consider $N$ total channel uses during the two time slots, then for AF TWRC, each time slot has $N/2$ channel uses. Let $P_{proc,1}$ be the average processing energy per channel use of the system, then its EE (bit/J) is
\begin{equation}
\xi=\frac{R_eN}{(P_A+P_B+P_R)N/2+P_{proc,1}N}.\label{EE,1}
\end{equation}
\indent
According to (\ref{total_energy}), the total processing energy $P_{proc,1}N$ can be divided into two parts: one is the processing energy in PA, which is linearly proportional to the transmission energy, the other is a constant value denoting energy dissipated in electronic circuits other than PA, i.e.,
\begin{equation}
P_{proc,1}N = (1/\eta-1)(P_A+P_B+P_R)N/2+P_{0,1}N, \label{proc1}
\end{equation}
where $(P_A+P_B+P_R)N/2$ is the transmission energy, and $P_{0,1}$ is the average circuit power consumption per channel use, except that consumed in PA. Substituting (\ref{BE1})(\ref{proc1}) into (\ref{EE,1}) gives
\begin{equation}
\xi=\frac{2R}{\frac{1}{\eta}(P_A+P_B+P_R)+2P_{0,1}}. \label{EE1}
\end{equation}
\subsection{Best Performance}
\begin{figure}
\centering
\includegraphics[scale=0.6]{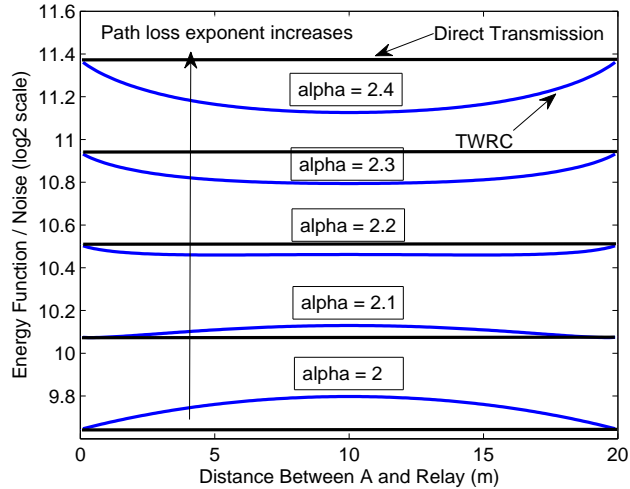}
\caption{Total energy consumption (normalized by $N_0$) for various locations of the relay, when $R = 1$\; bit/symbol, $\alpha$ varies from 2 to 2.4, $d = 20$\;m, $\eta = 1$, $P_{0,1} = 0$. Energy for direct transmission is plotted in straight lines.}\label{relay_location}
\end{figure}

\indent
We are now interested in finding the highest EE of AF TWRC when BE is given. Seeing from (\ref{BE1}) and (\ref{EE1}), given BE, EE is maximized when the denominator in (\ref{EE1}) is minimized. Define $f$ as the denominator of EE, i.e.,
\begin{equation}
f(R,h_A,h_B,\beta)= 1/\eta(P_A+P_B+P_R)+2P_{0,1}. \label{energy}
\end{equation}
$f$ equals the total energy consumption during two time slots, with each time slot occupying one channel use. Therefore, we call $f$ the energy function. For simplicity, let $h_1=|h_A|$, $h_2=|h_B|$, and $x=|\beta|^2$. Substituting (\ref{rate})(\ref{beta}) into (\ref{energy}) gives: $f(R,h_1,h_2,x)=$
\begin{equation}
\frac{N_0}{\eta}\Big(\frac{2(2^R-1)(1+h_1^2x)(1+h_2^2x)}{h_1^2h_2^2x}+x\Big)+2P_{0,1}. \label{energy_expand}
\end{equation}
Since $P_{0,1}$ is a constant, and $R$ is given by BE, there are three variables left, i.e., $h_1$, $h_2$, $x$. Thus, given BE, our goal is to find a vector ($x^*$,$h_1^*$,$h_2^*$) that offers the globally minimum value of $f$. \\
\indent
Let $d_1$, $d_2$ be the distance between A and relay, B and relay, respectively. For simplicity, we assume that the EM wave experiences large-scale fading only, i.e., $h_1^2=d_1^{-\alpha}$, $h_2^2=d_2^{-\alpha}$, where $\alpha$ is the path loss exponent. This is a common assumption when one wants to theoretically relate channel gain to distance (or location), e.g., \cite{full}~\cite{EE-BE}. Assuming that the direct distance between A and B is fixed, denoted by $d$. If relay is put too far away, e.g., $d_1>d$, then A and B would prefer direct transmission without relaying. Therefore, the variables $h_1$, $h_2$ and $x$ are restricted by the following conditions
\begin{equation}
\begin{split}x>0,\; h_1>0,\;h_2>0\\
h_1^{-2/\alpha}<d,\;h_2^{-2/\alpha}<d\\
h_1^{-2/\alpha}+h_2^{-2/\alpha}\ge d \end{split}. \label{condition}
\end{equation}
\indent
Taking partial derivative of $f$ with respect to $x$, and let it be zero gives
\begin{equation}
x_0=\sqrt{\frac{2(2^R-1)}{(2^{R+1}-1)h_1^2h_2^2}}, \label{x0}
\end{equation}
and $\partial f/\partial x <0$ for $x<x_0$, $\partial f/\partial x >0$ for $x>x_0$. Therefore, the global minimum point $(x^*,h_1^*,h_2^*)$ of $f$, if exists, must satisfy (\ref{x0}), for if not, we can always find a $x_0^*$ calculated from (\ref{x0}) so that $f(x_0^*,h_1^*,h_2^*)<f(x^*,h_1^*,h_2^*)$, which contradicts the assumption that $f$ achieves global minimum at $(x^*,h_1^*,h_2^*)$. Substituting (\ref{x0}) into (\ref{energy_expand}) gives
\begin{equation}
f(R,h_1,h_2,x_0)=\frac{ah_1h_2+b(h_1^2+h_2^2)}{\eta h_1^2h_2^2}+2P_{0,1}, \label{energy_x0}
\end{equation}
where
\begin{equation}
a=2N_0\sqrt{(2^{R+1}-1)(2^{R+1}-2)} \label{a}
\end{equation}
and
\begin{equation}
b=N_0(2^{R+1}-2).\label{b}
\end{equation}
\indent
Therefore, finding the minimum energy consumption is equal to finding the global minimum value of (\ref{energy_x0}). Since $\partial f/\partial h_1<0$ and $\partial f/\partial h_2<0$, given $h_1$ ($h_2$), $f$ decreases when $h_2$ ($h_1$) is increased. Then the minimization will be achieved when $h_1$, $h_2$ ($h_1^{-2/\alpha}$, $h_2^{-2/\alpha}$) are as large (small) as possible, so from (\ref{condition}), we have that $f$ achieves its minimum value when $h_1^{-2/\alpha}+h_2^{-2/\alpha}=d$. This means that TWRC consumes less energy when relay is located along the straight line connecting A and B, which agrees with our intuition. Let $h_1^{-2/\alpha}=d\cos^2\theta$, $h_2^{-2/\alpha}=d\sin^2\theta$, where $\theta\in(0,\pi/2)$, then (\ref{energy_x0}) becomes
\begin{equation}
f(\theta)=\frac{ad^{\alpha}\cos^{\alpha} \theta \sin^{\alpha}\theta+bd^{\alpha}(\sin^{2\alpha}\theta+\cos^{2\alpha}\theta)}{\eta} +2P_{0,1}.
\end{equation}
\indent
Its first derivative is zero when $\theta=\pi/4$, which corresponds to the situation when relay is in the middle between A and B. To see whether $f$ achieves local minimum at this critical value, we check its second derivative and get
\begin{equation}
f(\pi/4) \;\text{achieves} \begin{cases} \text{local min} & \alpha>1+\sqrt{\frac{2^{R+1}-1}{2^{R+1}-2}}\\
\text{local max} & \alpha<1+\sqrt{\frac{2^{R+1}-1}{2^{R+1}-2}}. \end{cases}\label{case}
\end{equation}
\begin{figure}
\centering
\includegraphics[scale=0.6]{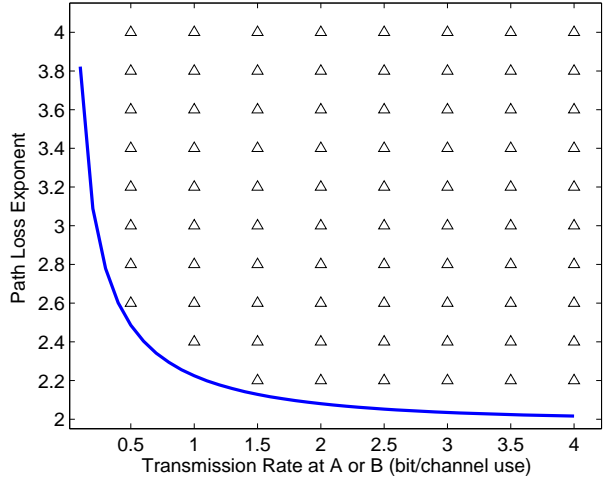}
\caption{The ($R$, $\alpha$) region where energy consumption is minimized when relay is in the middle.}\label{equi_dist}
\end{figure}
\indent
From (\ref{case}), we know that whether relay in the middle achieves local minimum energy consumption is determined by whether the path loss exponent is above the threshold formed by the transmission rate. Fig.~\ref{equi_dist} shows in what $(R, \alpha)$ region relay in the middle consumes locally minimum energy, from which we see that this proposition holds true in most cases. In $802.11a$ standards~\cite{802.11a}, for example, we have $54$ Mbps and $20$ MHz, so $R = 2.7 $ bit/channel use, then as long as $\alpha > 2.04$, transmission through TWRC with relay in the middle saves the most energy. Fig.~\ref{relay_location} plots the energy function divided by $N_0$ for both AF TWRC and direct transmission between A and B, assuming $R = 1$\; bit/symbol, $\alpha$ = 2 - 2.4, $d = 20$\;m, and $P_{proc,1} = 0$. The x-axis denotes a relay's location along the line connecting A and B. Note that for different $d$ and $P_{proc,1}$, the absolute value of each curve will be different, but here we are interested in their shape as well as the comparison results of their magnitudes, which will not change with $d$ and $P_{proc,1}$. In (\ref{case}), let $R = 1$\; bit/symbol, we get that when $\alpha > 2.22$, $f$ is locally minimized when relay is in the middle, which can be confirmed by Fig.~\ref{relay_location}. As shown in this figure, $f$ achieves not only local minimum but also global minimum under the condition $\alpha > 2.22$. When $\alpha \le 2.22$, $f$ is locally maximized when relay is in the middle, although not always globally maximized, as shown in Fig.~\ref{relay_location}. As $\alpha$ decreases to $2$, the energy consumption for TWRC tends to be higher than direct transmission, which means that direct transmission works better in terms of less energy dissipation. Together with (\ref{case}), we conclude that as long as $\alpha>1+\sqrt{\frac{2^{R+1}-1}{2^{R+1}-2}}$, resources are better utilized when transmitting through TWRCs, rather than direct transmission; and relay is best located at the middle point. Discussions in the following paper will be focused on this ($R$,$\alpha$) region.\\
\section{AF TWRC in A Wireless Network}
Previously, we studied a simple route with only one relay, and derived its optimal EE given BE. In this section, we consider routes with more than one relays. We first propose the network model. To analyze different routes in terms of EE, BE and latency, we need a scheduling scheme that enables transmission through AF TWRCs along a multihop route, so a simple scheme is presented. Based on this simple scheduling scheme, we then derive the performance measures of routes with different number of relays.
\subsection{Network Model}
\begin{figure}[ht]
\centering
\includegraphics[scale=0.6]{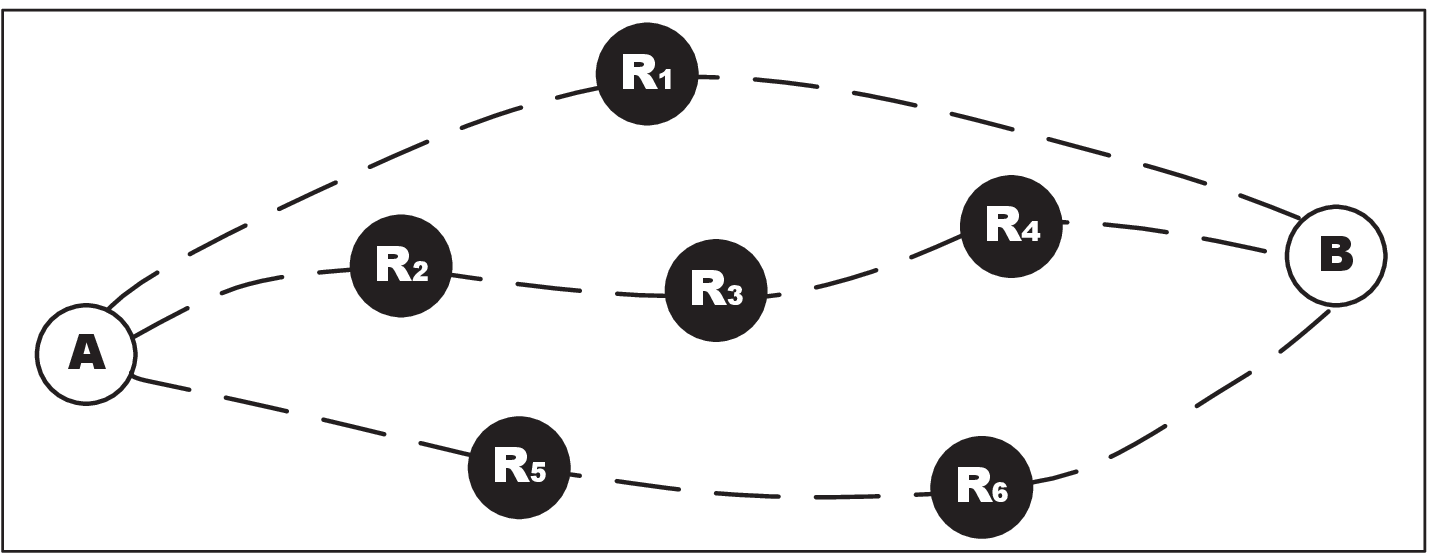}
\caption{A and B want to exchange data through AF TWRCs, which route performs best?}\label{model}
\end{figure}
A single-frequency AF TWRC network contains multiple pairs of nodes that want to exchange data through multiple relays. Perfect scheduling is assumed in the network so that each source-destination (SD) pair transmit in its own time slot with no concurrent transmission inside its interference range. Therefore, the whole network can be decomposed into multiple small networks each with only one SD pair. Moreover, considering the whole network is equivalent to considering each of the small networks separately. Thus, it's enough to analyze the routing path selection problem for one small network, i.e., one SD pair. Since our analysis is based on the assumption of perfect scheduling, the results gained will provide upper bound of the network performance.\\
\indent
We now focus on one SD pair. Suppose A and B want to exchange data through TWRCs, as shown in Fig.~\ref{model}. Among the several routes, each with different length and number of relays, we want to know which route gives the best performance. To better use the results in Section II, we make the following two assumptions:
\begin{itemize}
\item the relays along each route are equidistant;
\item each route has a small curvature so that it can be treated as an elongated straight line.
\end{itemize}
The two assumptions are reasonable. Practically, the relays will not be allocated too near or too far from each other. Besides, based on the analysis from previous section, in most cases, relay located in the middle consumes the smallest energy, and hence equally spaced relays along a route tends to achieve low energy consumption. Route with a large curvature means that most energy is spent for data in moving around the source node rather than forwarding to its destination. Besides, nodes with several hops in between may be close in distance so that they cannot transmit simultaneously, which decreases the concurrent number of transmission nodes. Therefore, route with a large curvature needs to be avoided.\\
\indent
In addition, let $k$ be the number of relays along a route, this paper considers the case when $0 \le k \le 6$, for TWRC is usually used in a small-scaled wireless access network, which is sure to have an upper limit on the number of hops. Note that our framework can be extended to the case when $k>7$, with a possibly increasing complexity as $k$ becomes larger. For simplicity, we also assume that each node works in half-duplex mode and that each hop transmits at a constant rate. This constant rate control mechanism ensures that each node forwards a packet to the next hop per time slot, which help maintain the stability of the system.
\subsection{Two Multi-hop Transmission Schemes}
\begin{figure}
\centering
\includegraphics[scale=0.6]{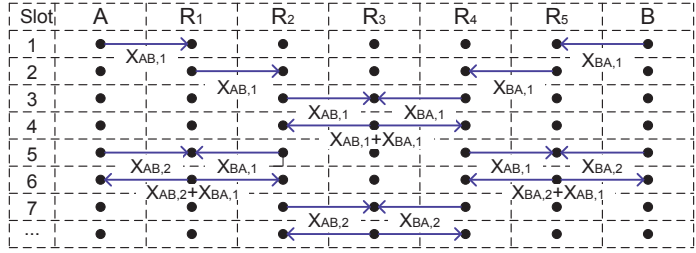}
\caption{Illustration of the Hop-by-Hop scheme for $k=5$.}\label{hop-by-hop}
\end{figure}

\begin{figure}[ht]
\centering
\includegraphics[scale=0.6]{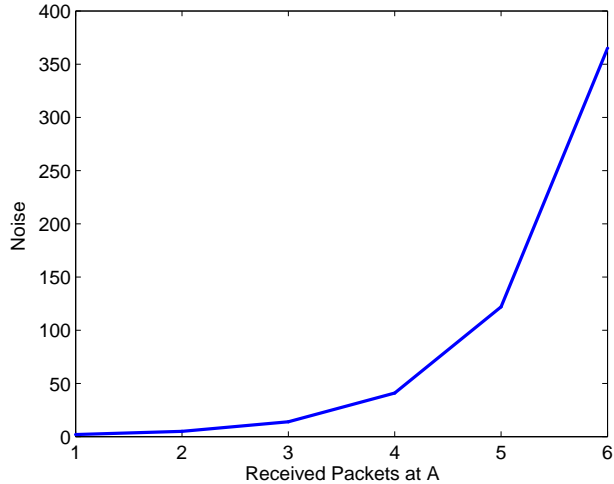}
\caption{In End-to-End scheme, noise received at the end nodes increases exponentially with the number of received packets.}\label{noise}
\end{figure}

\indent
To enable multi-hop transmission through TWRCs, two schemes can be used: Hop-by-Hop scheme and End-to-End scheme. We assume that the two end nodes always have packets to exchange. Taking $k = 5$ as an example, Fig.~\ref{hop-by-hop} illustrates how Hop-by-Hop scheme works. At the beginning, no packet is traversing along the route. Then the end nodes start to put in the packets that they want to exchange with each other. After two time slots, the system enters a stable state in which packets are transmitted in a recursive pattern: during every $4$ time slots, the end nodes will insert one new pair of packets into the system (i.e., packets that will be delivered to the other side), and receive one new pair of packets from the other side (i.e., packets that were inserted by the other side at an earlier time slot); all the relays in between will help forward data through TWRCs. In that way, the total time slots needed to exchange $n$ pairs of packets is $2+4n$, when $k = 5$. However, the recursive pattern is different when $k$ is different. As is illustrated in the next section, when $k$ is odd, then all the nodes are involved in TWRCs; when $k$ is even, then one of the end node will perform unicast transmission. Compared with Hop-by-Hop scheme, which is a loose ($4$ time slots / exchange) pattern formed by concatenated three-node subsystems and unicast channels, End-to-End scheme is a compact ($2$ time slots / exchange) pattern formed by overlapped three-node subsystems, where each node will interchange between transmitting and receiving every time slot.\\
\indent
It seems that End-to-End scheme is better since it is simpler and has higher throughput, however, this scheme experiences severe noise accumulation between adjacent three-node subsystems. As mentioned previously, End-to-End scheme has a compact transmission pattern, where only the two end nodes decode the received packets. Each relay in between performs subtraction, amplification, and broadcast. Correspondingly, the noisy packets will be subtracted from or superimposed with other noisy packets, resulting in a rapidly growing noise as it traverses the network. Fig.~\ref{noise} shows how noise accumulates at the end nodes as time elapses, assuming that the initial noise variance is $1$, relay does not do amplification, and noise accumulation due to substraction is ignored. The exponentially increased noise will terribly distort signals received at the end nodes. As to the Hop-by-Hop scheme, noise will not be accumulated between adjacent subsystems, since the two end nodes in each subsystem will decode the packet each time the relay broadcasts the overlapped waveforms. Therefore, although Hop-by-Hop scheme spends twice the time for the end nodes to exchange one packet, its performance is still better than End-to-End scheme.
\section{Performance of Hop-by-Hop Scheme}
\indent
In this section, we investigate the performance, i.e., BE, EE and latency for routes with different number of relays, based on the Hop-by-Hop scheme. Since how power is provisioned at each node affects the system's EE and BE, we will determine the optimal power allocation scheme under which a route achieves the highest EE with a given BE. The difficulty during derivation mainly comes from the interference influence. Since we are utilizing Shannon's capacity formula to derive the fundamental relation, any interference, no matter how small it is, should be included in the SINR part inside the formula, which increases complexity during the derivation process. Numerical analysis in the Section IV illustrates that ignoring this interference will cause big error percentage at high BE. Therefore, we need to carefully tackle with this interference influence.\\
\indent
Recall that relays are equally spaced, and that each node is transmitting at rate $R$ (bit/channel use) with capacity-achieving codes. Besides, each node operates in half-duplex mode. Denote by $h$ and $d$ the channel gain and the distance between two consecutive nodes (for simplicity, we use $h$ as both the channel gain and its Euclidean norm). Here, we only consider large-scale fading, i.e., $h^2 = d^{-\alpha}$. Let $P_i$ denote the transmission power per channel use at node $i$. We do not distinguish between $h$ and $P_i$ for different $k$, since there is no ambiguity in the following derivation. Let $P_{proc,k}$ be the average processing energy per channel use for route with $k$ relays, then Section II gives $P_{proc,k}=(1/\eta-1)P_{tr,k}+P_{0,k}$. Assuming that each channel use occupies $1$ second $\times$ hertz. Besides, let transmission range be $1$ hop, interference range be $2$ hops, i.e., nodes within two hops cannot transmit simultaneously, except in TWRCs.\\
\subsection{Latency}
In Hop-by-Hop scheme, packet is forwarded to the next hop per time slot, so the latency of a route with $k$ relays is $k+1$ time slot/bit.
\subsection{k=0}
\begin{figure}
\centering
\includegraphics[scale=0.6]{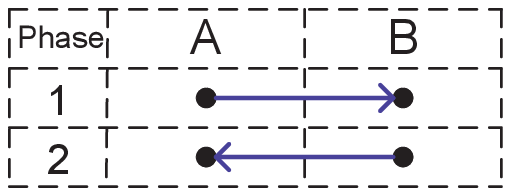}
\caption{Illustration of the Hop-by-Hop scheme: $k=0$.}\label{k=0}
\end{figure}
Fig.~\ref{k=0} illustrates $k=0$, i.e., direct transmission.
\begin{itemize}
\item \emph{BE and EE}
\end{itemize}
\begin{equation}
\gamma=R,\;\;
\xi=\frac{2R}{u(R,h)+2P_{proc,0}}.
\end{equation}
$u(R,h)=P_A+P_B$ is the transmission energy during the two time slots, with each time slot occupying one channel use. Besides, $2P_{proc,0}=(1/\eta-1)u(R,h)+2P_{0,0}$.
\begin{itemize}
\item \emph{Optimal Power Allocation}
\end{itemize}

Given BE, $u(R,h)$ achieves its minimal value when using capacity-achieving codes, i.e.,
\begin{equation}
R=\log_2(1+\frac{h^2P_i}{N_0}), \;i=\text{A or B}.
\end{equation}
Thus, the optimal power is
\begin{equation}
P_A=P_B=N_0(2^R-1)h^{-2}. \label{power,k=0}
\end{equation}
\subsection{k=1}
Fig.~\ref{TWRC} depicts the case when $k=1$, i.e., three-node TWRC. As analyzed in Section II, A and B exchange one packet every two time slots.
\begin{itemize}
\item \emph{BE and EE}
\end{itemize}
\begin{equation}
\gamma=R,\;\;
\xi=\frac{2R}{g(R,h)+2P_{proc,1}}. \label{EE,k=1}
\end{equation}
$g(R,h)=P_A+P_{R_1}+P_B$ is the transmission energy during the two time slots, with each time slot occupying one channel use. $2P_{proc,1}=(1/\eta-1)g(R,h)+2P_{0,1}$.
\begin{itemize}
\item \emph{Optimal Power Allocation}
\end{itemize}

Given BE, the minimal value of $g(R,h)$ is $g(R,h)_{\min}=(a+2b)h^{-2}$, which can be derived from (\ref{energy_x0}) by setting $h_1=h_2=h$, $\eta=1$, and $P_{0,1}=0$, with $a$ and $b$ given in (\ref{a})(\ref{b}). The corresponding optimal power allocation can be found by setting $h_A=h_B=h$ in ($\ref{rate}$)($\ref{beta}$), i.e.,
\begin{equation}
\begin{split}
P_A = P_B &= (2^R-1)(h^2\beta^2+1)N_0/(h^4\beta^2),\\
 &P_R=(2h^2P_A+N_0)\beta^2. \end{split} \label{power,k=1}
\end{equation}
The optimal $\beta^2$ is given in (\ref{x0}).
\subsection{k=2}
\begin{figure}
\centering
\includegraphics[scale=0.6]{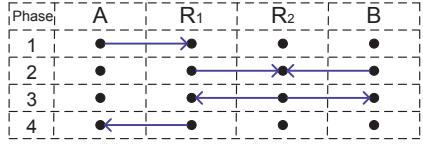}
\caption{Illustration of the Hop-by-Hop scheme: $k=2$.}\label{k=2}
\end{figure}

Fig.~\ref{k=2} illustrates how Hop-by-Hop scheme works when $k=2$. During the first and fourth time slots, A sends and receives one packet from R$_1$ through unicast transmission. During the second and third time slots, B sends and receives one packet from R$_2$ through TWRC. In this way, they form a recursive pattern of $4$ time slot/exchange.
\begin{itemize}
\item \emph{BE and EE}
\end{itemize}
\begin{equation}
\gamma=R/2,\;\;
\xi=\frac{2R}{u(R,h)+g(R,h)+4P_{proc,2}}.
\end{equation}
$u(R,h)=P_A+P_{R_1}$ and $g(R,h)=P_{R_1}+P_{R_2}+P_B$ represent transmission energy through unicast channel and TWRC, respectively. $4P_{proc,2}=(1/\eta-1)(u(R,h)+g(R,h))+4P_{0,2}$.
\begin{itemize}
\item \emph{Optimal Power Allocation}
\end{itemize}

Given EE, the minimum of $u(R,h)$ and $g(R,h)$ have been found in previous cases of $k=0$ and $k=1$, with the optimal power provision given by (\ref{power,k=0})(\ref{power,k=1}).
\subsection{k=3}
\begin{figure}
\centering
\includegraphics[scale=0.6]{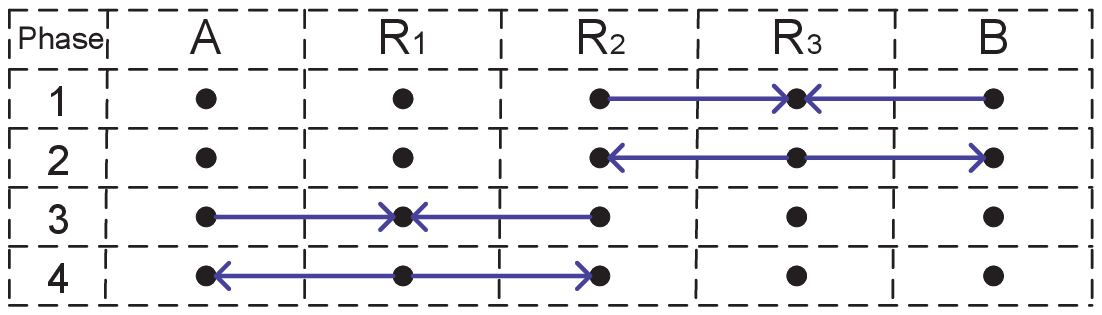}
\caption{Illustration of the Hop-by-Hop scheme: $k=3$.}\label{k=3}
\end{figure}
\begin{figure}
\centering
\includegraphics[scale=0.6]{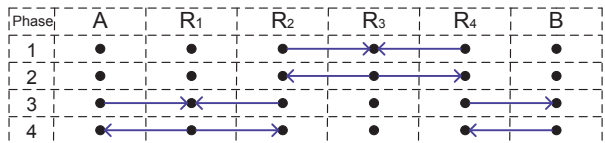}
\caption{Illustration of the Hop-by-Hop scheme: $k=4$.}\label{k=4}
\end{figure}
Fig.~\ref{k=3} presents the Hop-by-Hop scheme for $k=3$, where the left half nodes and the right half nodes form two TWRCs, respectively.
\begin{itemize}
\item \emph{BE and EE}
\end{itemize}
\begin{equation}
\gamma=R/2,\;\;
\xi=\frac{2R}{g_1(R,h)+g_2(R,h)+4P_{proc,3}}.
\end{equation}
$g_1(R,h)=P_A+P_{R_1}+P_{R_2}$ and $g_2(R,h)=P_{R_2}+P_{R_3}+P_{B}$ represent the transmission energy for the two TWRCs.
$4P_{proc,3}=(1/\eta-1)(g_1(R,h)+g_2(R,h)+4P_{0,3}$.
\begin{itemize}
\item \emph{Optimal Power Allocation}
\end{itemize}

Given EE, the minimal values of $g_1(R,h)$ and $g_2(R,h)$ are the same, as given in previous case of $k=1$, with the optimal power allocation in (\ref{power,k=1}).
\subsection{k=4}
Fig.~\ref{k=4} depicts the case for $k=4$. During the first two time slots, data is exchanged between R$_2$ and R$_4$. In the third time slot, A inserts a new packet into the system while B receives a packet. After that, the role of A and B exchanges in the fourth time slot. Thus, a recursive pattern is built with $4$ time slot/exchange.
\begin{itemize}
\item \emph{BE and EE}
\end{itemize}
\begin{equation}
\gamma=R/2,\;\;
\xi=\frac{2R}{g(R,h)+t(R,h)+4P_{proc,4}}.
\end{equation}
$g(R,h)=P_{R_2}+P_{R_3}+P_{R_4}$, $t(R,h)=P_{A}+P_{R_1}+P_{R_2}+P_{R_4}+P_{B}$ are the transmission energy consumed during the first two and last two time slots, respectively. $4P_{proc,4}=(1/\eta-1)(g(R,h)+t(R,h))+4P_{0,4}$.
\begin{itemize}
\item \emph{Optimal Power Allocation}
\end{itemize}

Given EE, the minimal value of $g(R,h)$ have been derived in the case of $k=1$, where $P_{R_2}$, $P_{R_3}$ and $P_{R_4}$ are given in (\ref{power,k=1}). Now we will find the minimal value of $t(R,h)$. A, R$_1$, R$_2$ form a TWRC, i.e.,
\begin{equation}
y_{R_1}=hx_{A}+hx_{R_2}+z_{R_1}+\sqrt{3^{-\alpha}}hx_{R_4},
\end{equation}
\begin{equation}
x_{R_1}=\beta y_{R_1}, \label{P_R1}
\end{equation}
\begin{equation}
y_{A}=hx_{R_1}+z_{A}+\sqrt{5^{-\alpha}}hx_{B},
\end{equation}
\begin{equation}
y_{R_2}=hx_{R_1}+z_{R_2}+\sqrt{3^{-\alpha}}hx_{B}.
\end{equation}
B, R$_4$ perform unicast transmission, i.e.,
\begin{equation}
y_{B}=hx_{R_4}+z_{B}+\sqrt{3^{-\alpha}}hx_{R_2}+\sqrt{5^{-\alpha}}hx_{A},
\end{equation}
\begin{equation}
y_{R_4}=hx_{B}+z_{R_4}+\sqrt{3^{-\alpha}}hx_{R_1}.
\end{equation}
If capacity-achieving codes are used, then
\begin{equation}
\frac{h^4\beta^2P_{R_2}}{3^{-\alpha} h^4\beta^2P_{R_4}+(h^2\beta^2+1)N_0+5^{-\alpha}h^2P_{B}}=2^R-1, \label{five_1}
\end{equation}
\begin{equation}
\frac{h^4\beta^2P_{A}}{3^{-\alpha}h^4\beta^2P_{R_4}+(h^2\beta^2+1)N_0+3^{-\alpha}h^2P_{B}}=2^R-1,\label{five_2}
\end{equation}
\begin{equation}
\frac{h^2P_{R_4}}{3^{-\alpha}h^2P_{R_2}+5^{-\alpha}h^2P_{A}+N_0}=2^R-1, \label{five_3}
\end{equation}
\begin{equation}
\frac{h^2P_{B}}{3^{-\alpha}h^2P_{R_1}+N_0}=2^R-1. \label{five_4}
\end{equation}
From (\ref{P_R1}) we also have
\begin{equation}
P_{R_1}=\beta^2(h^2P_{A}+h^2P_{R_2}+3^{-\alpha}h^2P_{R_4}+N_0). \label{five_5}
\end{equation}
There are $5$ equations, i.e., (\ref{five_1})-(\ref{five_5}), and $5$ variables, i.e., $P_{A}$, $P_{R_1}$, $P_{R_2}$, $P_{R_4}$, $P_{B}$. Thus, we are able to solve the equations and get the $5$ variables. \\
(\ref{five_4})(\ref{five_5}) gives
\begin{equation}
P_{R_1}=\frac{3^{\alpha}P_{B}}{2^R-1}-3^{\alpha}h^{-2}N_0. \label{five_7}
\end{equation}
Substituting (\ref{five_7}) and (\ref{five_3}) into the formula of $t(R,h)$ gives
\begin{equation}
t=(1+(2^R-1)5^{-\alpha})P_A+(1+(2^R-1)3^{-\alpha})P_{R_2}+(\frac{3^{\alpha}}{2^R-1}+1)P_B+((2^R-1)-3^{\alpha})h^{-2}N_0. \label{five_8}
\end{equation}
Eliminating $P_{R_4}$ from (\ref{five_1}) and (\ref{five_4}) gives
\begin{equation}
2^Rh^4\beta^2P_{R_2}+h^4\beta^2(2^R-1)P_{A}=(3^{\alpha}h^2+5^{-\alpha}h^2(2^R-1))P_{B}+(2^R-1)(1-3^{\alpha})N_0. \label{five_9}
\end{equation}
Eliminating $P_{R_4}$ from (\ref{five_2}) and (\ref{five_4}) gives
\begin{equation}
2^Rh^4\beta^2P_{A}+h^4\beta^2(2^R-1)P_{R_2}=(3^{\alpha}h^2+3^{-\alpha}h^2(2^R-1))P_{B}+(2^R-1)(1-3^{\alpha})N_0. \label{five_10}
\end{equation}
From (\ref{five_9}) and (\ref{five_10}), we get
\begin{equation}
(2^{R+1}-1)h^4\beta^2P_{A}=Bh^2P_{B}+(2^R-1)(1-3^{\alpha})N_0, \label{five_11}
\end{equation}
\begin{equation}
(2^{R+1}-1)h^4\beta^2P_{R_2}=a_2h^2P_{B}+(2^R-1)(1-3^{\alpha})N_0. \label{five_12}
\end{equation}
Substituting (\ref{five_3}) into (\ref{five_4}), together with (\ref{five_11}) and (\ref{five_12}), we get
\begin{equation}
(1-a_5)h^2P_{B}=(a_3+h^2\beta^2a_4)N_0. \label{five_13}
\end{equation}
Substituting (\ref{five_11})(\ref{five_12})(\ref{five_13}) into the formula of (\ref{five_8}) gives
\begin{equation}
t(R,h)=(a_8\beta^2+\frac{a_7}{h^4\beta^2}+a_6h^{-2})N_0\ge (2\sqrt{a_7a_8}+a_6)N_0h^{-2}. \label{t_min}
\end{equation}
So the minimum of $t$ is $t(R,h)_{\min}=(2\sqrt{a_7a_8}+a_6)N_0h^{-2}$, achieved when $\beta^4=a_7h^{-4}/a_8$. The corresponding optimal power allocation is
\begin{equation}
P_B= \frac{(a_3+a_4h^2\beta^2)N_0}{(1-a_5)h^2},
\end{equation}
\begin{equation}
P_A=\frac{a_1h^2P_B+(2^R-1)(1-3^{\alpha})N_0}{(2^{R+1}-1)h^4\beta^2},
\end{equation}
\begin{equation}
P_{R_2} =\frac{a_2h^2P_B+(2^R-1)(1-3^{\alpha})N_0}{(2^{R+1}-1)h^4\beta^2}.
\end{equation}
The optimal power allocated at R$_1$ and R$_4$ can then be easily computed from (\ref{five_3})(\ref{five_4}). $\{a_i\}_{i=1}^{8}$ are functions of $R$ and $\alpha$:
\begin{equation}
a_1=-(2^R-1)^25^{-\alpha}+2^R(2^R-1)3^{-\alpha}+3^{\alpha},
\end{equation}
\begin{equation}
a_2=2^R(2^R-1)5^{-\alpha}-(2^R-1)^23^{-\alpha}+3^{\alpha},
\end{equation}
\begin{equation}
a_3=\frac{(2^R-1)^2}{2^{R+1}-1}(1-3^{\alpha})(3^{-\alpha}2+(45^{-\alpha}+27^{-\alpha})(2^R-1))+(2^R-1),
\end{equation}
\begin{equation}
a_4=9^{-\alpha}(2^R-1)^2+3^{-\alpha}(2^R-1),
\end{equation}
\begin{equation}
a_5=\frac{2^R-1}{2^{R+1}-1}[a_2(3^{-\alpha}+45^{-\alpha}(2^R-1))+a_1(3^{-\alpha}+27^{-\alpha}(2^R-1))],
\end{equation}
\begin{equation}
a_6=\frac{a_2a_4(1+(2^R-1)5^{-\alpha})+a_1a_4(1+(2^R-1)3^{-\alpha})}{(2^{R+1}-1)(1-a_5)}+(2^R-1)-3^{\alpha}+\frac{a_3}{1-a_5}(\frac{3^{\alpha}}{2^R-1}+1),
\end{equation}
\begin{equation}
a_7=\frac{(2^R-1)(1-3^{\alpha})}{2^{R+1}-1}(2+(2^R-1)(3^{-\alpha}+5^{-\alpha}))+\frac{a_2a_3(1+(2^R-1)5^{-\alpha})}{(1-a_5)(2^{R+1}-1)}
+\frac{a_1a_3(1+(2^R-1)3^{-\alpha})}{(1-a_5)(2^{R+1}-1)},
\end{equation}
\begin{equation}
a_8=\frac{a_4}{1-a_5}(\frac{3^{\alpha}}{2^R-1}+1).
\end{equation}
\subsection{k=5}
\begin{figure}
\centering
\includegraphics[scale=0.6]{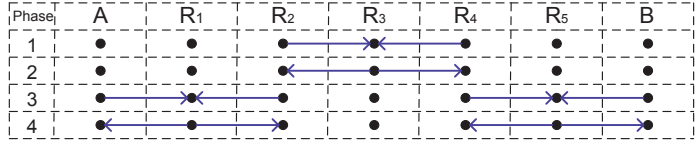}
\caption{Illustration of the Hop-by-Hop scheme: $k=5$.}\label{k=5}
\end{figure}
Fig.~\ref{k=5} illustrates how Hop-by-Hop scheme works when $k=5$. This recursive pattern has been analyzed in previous subsection.
\begin{itemize}
\item \emph{BE and EE}
\end{itemize}
\begin{equation}
\gamma=R/2,\;\;
\xi=\frac{2R}{g(R,h)+s(R,h)+4P_{proc,5}}.
\end{equation}
$g(R,h)=P_{R_2}+P_{R_3}+P_{R_4}$, $s(R,h)=P_{A}+P_{R_1}+P_{R_2}+P_{R_4}+P_{R_5}+P_{B}$ are the transmission energy consumed during the first two and last two time slots, respectively. $4P_{proc,5}=(1/\eta-1)(g(R,h)+s(R,h))+4P_{0,5}$.
\begin{itemize}
\item \emph{Optimal Power Allocation}
\end{itemize}

Similar to the previous case, given EE, the minimal value of $g(R,h)$ has been derived when $k=1$. Next we find the minimal value of $s(R,h)$ and the corresponding power allocation. A, R$_1$, R$_2$, and B, R$_4$, R$_5$ form two TWRCs, i.e.,
\begin{equation}
y_{R_1}=hx_{A}+hx_{R_2}+\sqrt{3^{-\alpha}}hx_{R_4}+\sqrt{5^{-\alpha}}hx_{B}+z_{R_1},
\end{equation}
\begin{equation}
x_{R_1}=\beta y_{R_1}, \label{amp}
\end{equation}
\begin{equation}
x_{A}=hx_{R_1}+\sqrt{5^{-\alpha}}hx_{R_5}+z_{A},
\end{equation}
\begin{equation}
x_{R_2}=hx_{R_1}+\sqrt{3^{-\alpha}}hx_{R_5}+z_{R_2}.
\end{equation}
Assuming capacity-achieving codes, then
\begin{equation}
\frac{h^4\beta^2P_{R_2}}{h^2(P_{R_1}-\beta^2h^2(P_A+P_{R_2}))+5^{-\alpha}h^2P_{R_5}+N_0}=2^R-1, \label{three_1}
\end{equation}
\begin{equation}
\frac{h^4\beta^2P_{A}}{h^2(P_{R_1}-\beta^2h^2(P_A+P_{R_2}))+3^{-\alpha}h^2P_{R_5}+N_0}=2^R-1. \label{three_2}
\end{equation}
From (\ref{amp}), we also have
\begin{equation}
P_{R_1}=\beta^2(h^2P_A+h^2P_{R_2}+3^{-\alpha}h^2P_{R_4}+5^{-\alpha}h^2P_B+N_0).\label{three_3}
\end{equation}
Due to symmetry, $P_A=P_B$, $P_{R_2}=P_{R_4}$, $P_{R_1}=P_{R_5}$, so we only need to consider the one of the two TWRCs. Accordingly, $s(R,h)=2(P_A+P_{R_1}+P_{R_2})$.
Similar to the case of $k=4$, we can solve the $3$ equations, i.e., (\ref{three_1})-(\ref{three_3}).\\
Substituting (\ref{three_3}) into the formula of $s(R,h)$ gives
\begin{equation}
s=2(1+(1+5^{-\alpha})h^2\beta^2)P_A+2(1+(1+3^{-\alpha})h^2\beta^2)P_{R_2}+2\beta^2N_0. \label{three_5}
\end{equation}
(\ref{three_2}) minus (\ref{three_1}) gives
\begin{equation}
h^4\beta^2(P_{A}-P_{R_2})=(2^R-1)(3^{-\alpha}-5^{-\alpha})h^2P_{R_1}. \label{three_6}
\end{equation}
Substituting (\ref{three_3}) into (\ref{three_6}) gives
\begin{equation}
h^4\beta^2P_A=b_1h^4\beta^2P_{R_2}+b_2h^2\beta^2N_0. \label{three_7}
\end{equation}
Substituting (\ref{three_7}) into (\ref{three_1}) and (\ref{three_2}) gives
\begin{equation}
h^4\beta^2P_{R_2}=(b_3h^2\beta^2+b_4)N_0. \label{three_8}
\end{equation}
\begin{equation}
h^4\beta^2P_A=((b_1b_3+b_2)h^2\beta^2+b_1b_4)N_0. \label{three_9}
\end{equation}
Substituting (\ref{three_8}) and (\ref{three_9}) into (\ref{three_5}) gives
\begin{equation}
s(R,h)=(b_5\beta^2+\frac{b_6}{h^4\beta^2}+b_7h^{-2})N_0\ge (2\sqrt{b_5b_6}+b_7)N_0h^{-2}. \label{s_min}
\end{equation}
So the minimum of $s$ is $s(R,h)_{\min}=(2\sqrt{b_5b_6}+b_7)N_0h^{-2}$, achieved when $\beta^4=b_6h^{-4}/b_5$. The corresponding optimal power allocation is
\begin{equation}
P_A = \frac{((b_1b_3+b_2)h^2\beta^2+b_1b_4)N_0}{h^4\beta^2},
\end{equation}
\begin{equation}
P_{R_2} = \frac{(b_3h^2\beta^2+b_4)N_0}{h^4\beta^2}.
\end{equation}
The optimal power allocated at R$_1$ can be easily computed from (\ref{three_3}). $\{b_i\}_{i=1}^{7}$ are complicated functions of $R$ and $\alpha$, and their full expressions are
\begin{equation}
b_1=\frac{1+(2^R-1)(3^{-\alpha}-5^{-\alpha})(1+3^{-\alpha})}{1-(2^R-1)(3^{-\alpha}-5^{-\alpha})(1+5^{-\alpha})},
\end{equation}
\begin{equation}
b_2=\frac{(2^R-1)(3^{-\alpha}-5^{-\alpha})}{1-(2^R-1)(3^{-\alpha}-5^{-\alpha})(1+5^{-\alpha})},
\end{equation}
\begin{equation}
b_3=\frac{[(5^{-\alpha}2+25^{-\alpha})b_2+5^{-\alpha}+1](2^R-1)}{1-(2^R-1)[b_15^{-\alpha}(2+5^{-\alpha})+5^{-\alpha}+3^{-\alpha}+15^{-\alpha}]},
\end{equation}
\begin{equation}
b_4=\frac{2^R-1}{1-(2^R-1)[b_15^{-\alpha}(2+5^{-\alpha})+5^{-\alpha}+3^{-\alpha}+15^{-\alpha}]},
\end{equation}
\begin{equation}
b_5=2(b_1b_3+b_2)(1+5^{-\alpha})+2b_3(1+3^{-\alpha})+2
\end{equation}
\begin{equation}
b_6=2(b_1+1)b_4
\end{equation}
\begin{equation}
b_7=2(b_1b_3+b_2)+2b_1b_4(1+5^{-\alpha})+2b_3+2b_4(1+3^{-\alpha}).
\end{equation}
\subsection{k=6}
The recursive pattern of $k=6$ resembles that of $k=4$ and $k=5$. During the first two time slots, the system works like in the third and fourth time slots when $k=4$. During the last two time slots, the recursive pattern is the same as that when $k=5$.
\begin{itemize}
\item \emph{BE and EE}
\end{itemize}
\begin{equation}
\gamma=R/2,\;\;
\xi=\frac{2R}{t(R,h)+s(R,h)+4P_{proc,6}}.
\end{equation}
$t(R,h)=P_{R_2}+P_{R_3}+P_{R_4}+P_{R_6}+P_{B}$ and $s(R,h)= P_{A}+P_{R_1}+P_{R_2}+P_{R_4}+P_{R_5}+P_{R_6}$ give the transmission energy during the first and the last two time slots, respectively. $4P_{proc,6}=(1/\eta-1)(t(R,h)+s(R,h))+4P_{0,6}$.
\begin{itemize}
\item \emph{Optimal Power Allocation}
\end{itemize}

Due to the same recursive pattern, the minimal values of $t(R,h)$ and $s(R,h)$ as well as the optimal power allocation have been derived in previous cases of $k=4$ and $k=5$.
\section{Numerical Results}
\begin{figure}
\centering
\subfigure[]{\label{total_large_proc}
\includegraphics[scale=0.5]{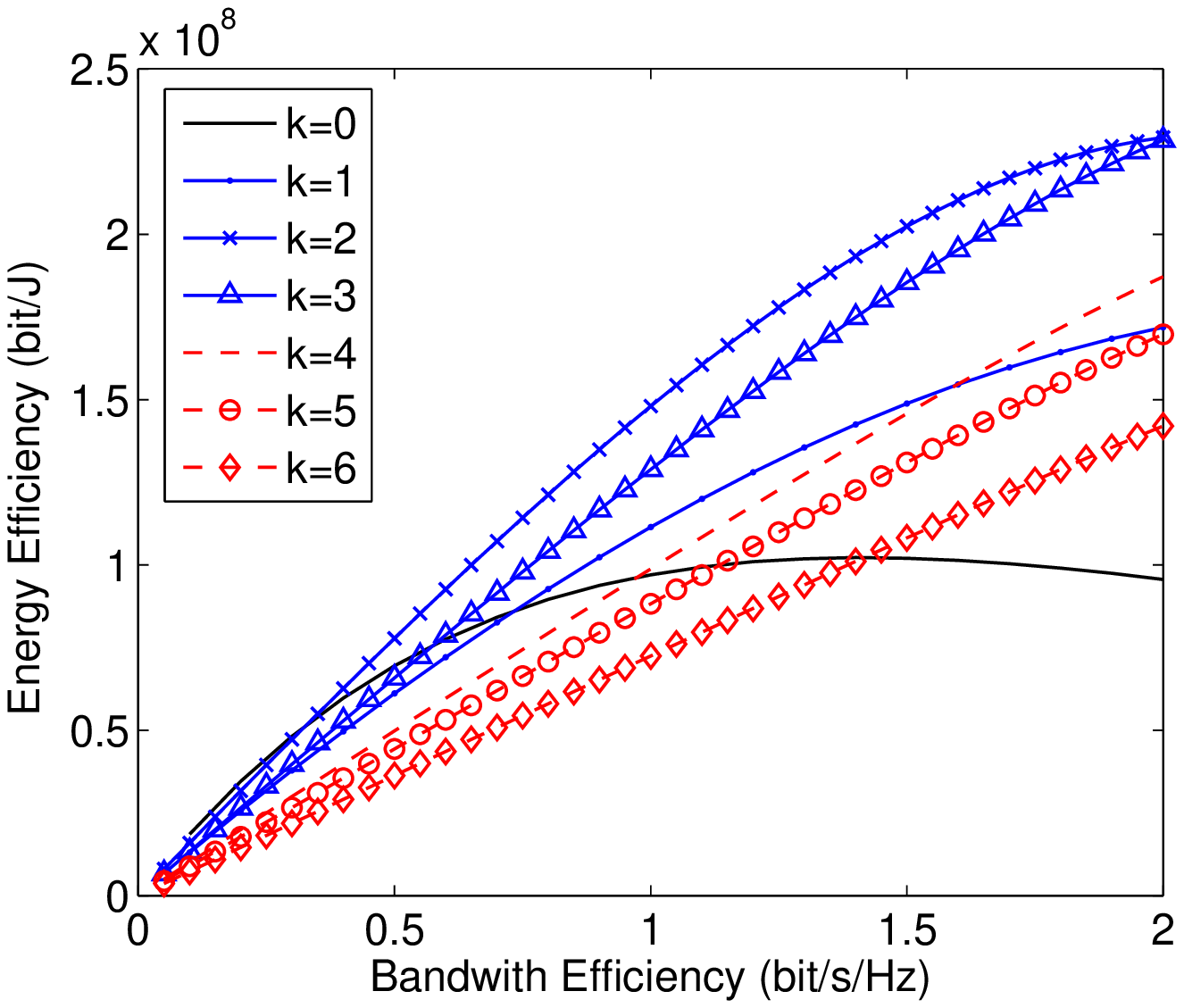}}
\subfigure[]{\label{total_small_proc}
\includegraphics[scale=0.5]{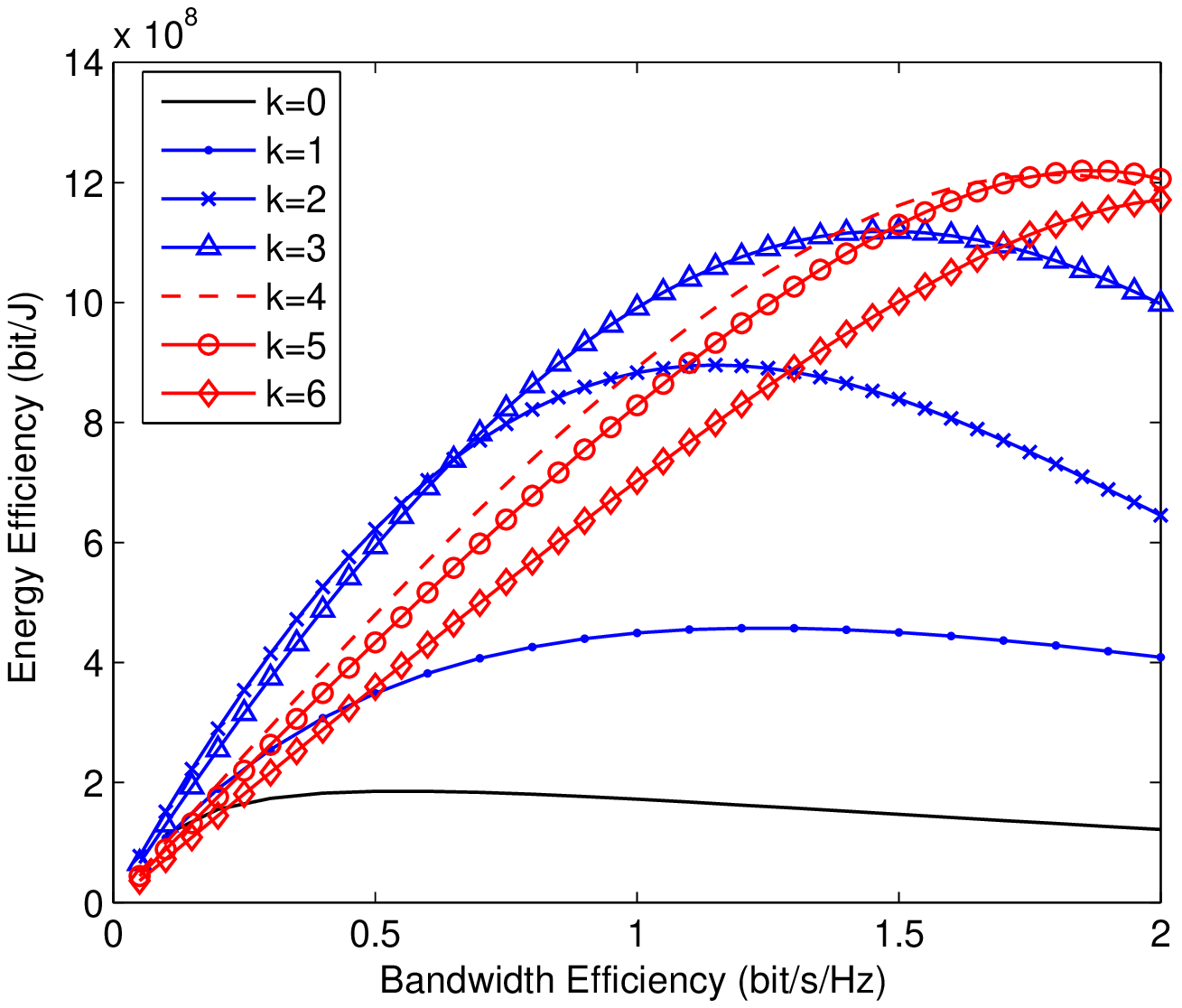}}
\caption{EE-BE relation for different number of relays ($k$ varies from $0$ to $6$) when $d_{route}= 1000$ m, $\alpha=4$, $\eta=0.75$, (a) $P_{0,0}=5\times 10^{-6}$ (b) $P_{0,0}=5\times 10^{-7}$.}
\end{figure}

\indent
In Section VI, we have derived the performance measures of routes with different number of relays. We also found the optimal power allocation scheme, under which the route consumes the smallest energy while still transmits at the same end-to-end rate. In this section, we numerically study the performance of different routes with the application of the optimal power allocation scheme.\\
\indent
Let path loss exponent $\alpha=4$, then from (\ref{case}) we know that only when the transmission rate $R>0.087$ bits/channel use, does the relay in the middle minimize the whole energy dissipation. Besides, set noise variance as $N_0=-174$ dBm/Hz, and drain efficiency $\eta=0.75$, which is an achievable value for high-class power amplifiers~\cite{processing}. For the processing energy, we assume that $P_{0,k}$ is proportional to the average number of senders and receivers per channel use, i.e., $P_{0,k}/P_{0,0}=m/2$, where $m$ is the average number of active nodes per channel use for route with $k$ relays. Let $P_{0,0}=5\times 10^{-7}$ mJ/channel use, then $P_{0,1}=3/2P_{0,0}$, $P_{0,2}=5/4P_{0,0}$, $P_{0,3}=7/4P_{0,0}$, $P_{0,4}=2P_{0,0}$, $P_{0,5}=9/4P_{0,0}$, $P_{0,6}=11/4P_{0,0}$. $d_{route}$ is the total length of the route.\\
\begin{figure}
\centering
\includegraphics[scale=0.6]{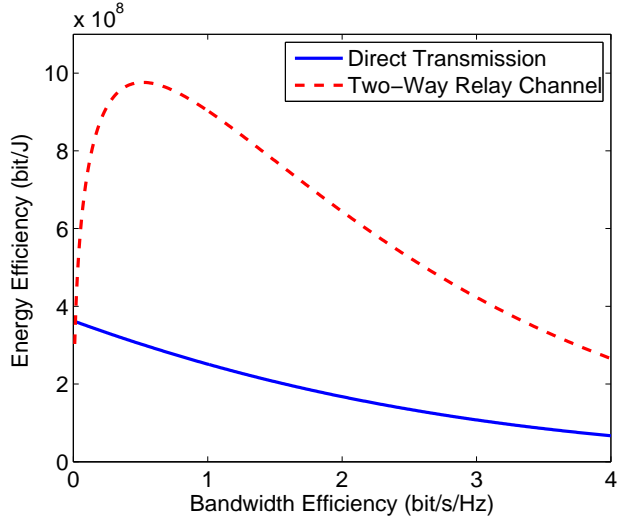}
\caption{Comparison of the EE-BE relation for direct transmission ($k=0$) and TWRC ($k=1$) when end-to-end distance $d=1000$ m, $\alpha=4$, and processing energy is ignored.}\label{direct_vs_twrc}
\end{figure}
\indent
Fig.~\ref{direct_vs_twrc} compares the performance for direct transmission and TWRC, which corresponds to $k=0$ and $k=1$. Since both of them spend $2$ time slots in exchanging one pair of packets, they have the same latency, so only EE and BE are compared in the figure. Here we ignore the processing energy, since nonzero processing energy will not influence the results. For direct transmission, EE and BE are always negatively related, which is a fundamental feature inherited from the Shannon¡¯s capacity formula~\cite{verdu}. However, TWRC changes the way that EE-BE interacts: EE and BE are positively related at low rate; and then involved in a tradeoff relation after EE reaches its maximum value. In other words, at low rate, we can decrease energy consumption and increase transmission rate at the same time. This feature can also be seen from the equation (\ref{EE,k=1}). If processing energy is ignored, then EE for a three-node TWRC is
\begin{equation}
\xi=\frac{R}{N_0(\sqrt{(2^{R+1}-1)(2^{R+1}-2)}+2^{R+1}-2)h^{-2}}.
\end{equation}
It can be verified that $\lim_{R\to 0}\xi=0$ and $\lim_{R\to \infty}\xi=0$, there must be a transmission rate that achieves maximum EE. From Fig.~\ref{direct_vs_twrc}, we can estimate that this rate is $R\approx 0.6$ bit/channel use. Besides, from the figure, we see that when rate approaches zero, direct transmission tends to have higher EE than TWRC, which is consistent with analysis in Section II. Therefore, in the following analysis, we focus on $R>0.087$ bit/channel use, where TWRC will have better performance than direct transmission, and energy consumption is minimized when relay is in the middle.\\
\begin{figure}
\centering
\subfigure[]{\label{txpower_k06}
\includegraphics[scale=0.5]{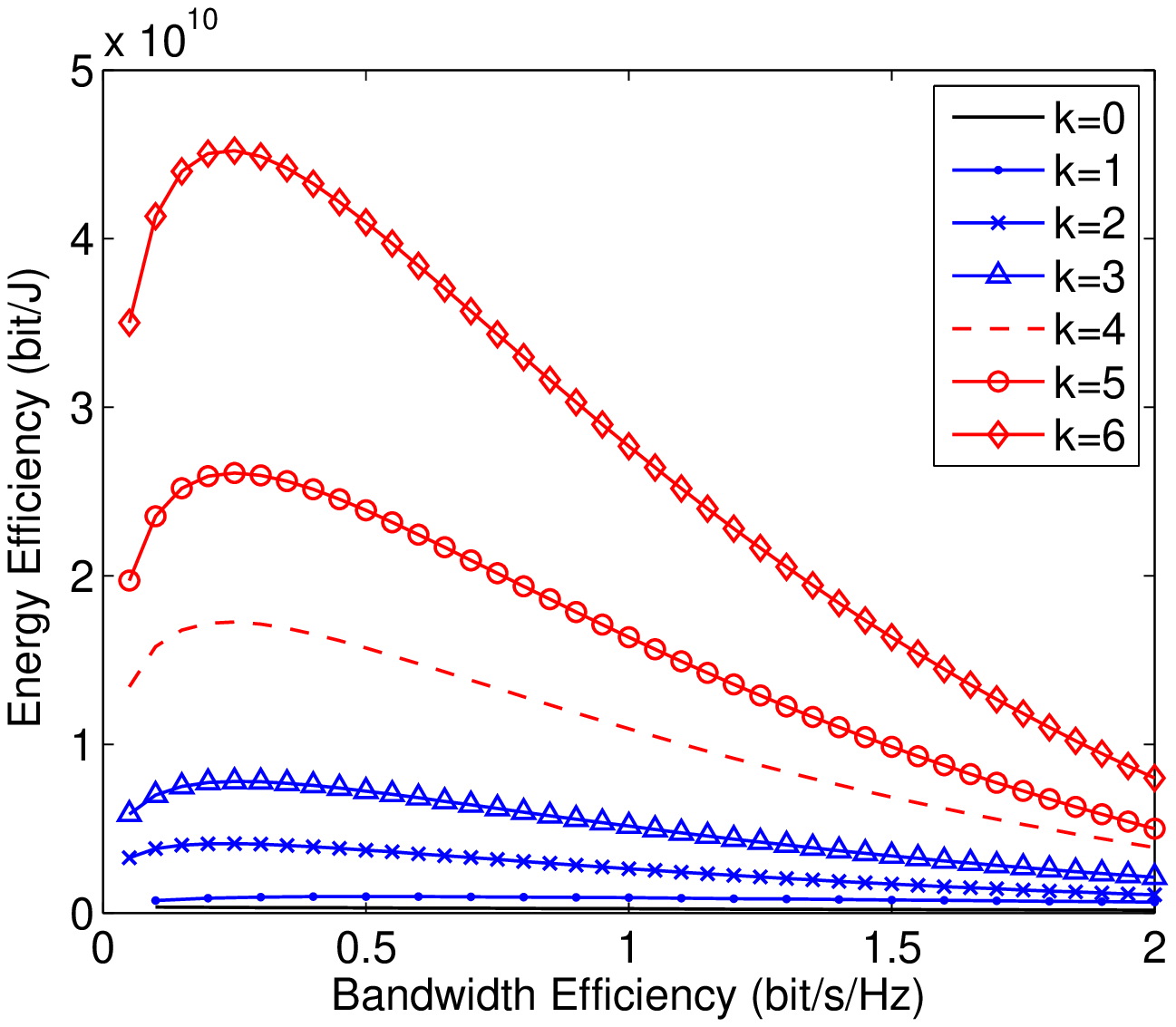}}
\subfigure[]{\label{txpower_compa_k36}
\includegraphics[scale=0.5]{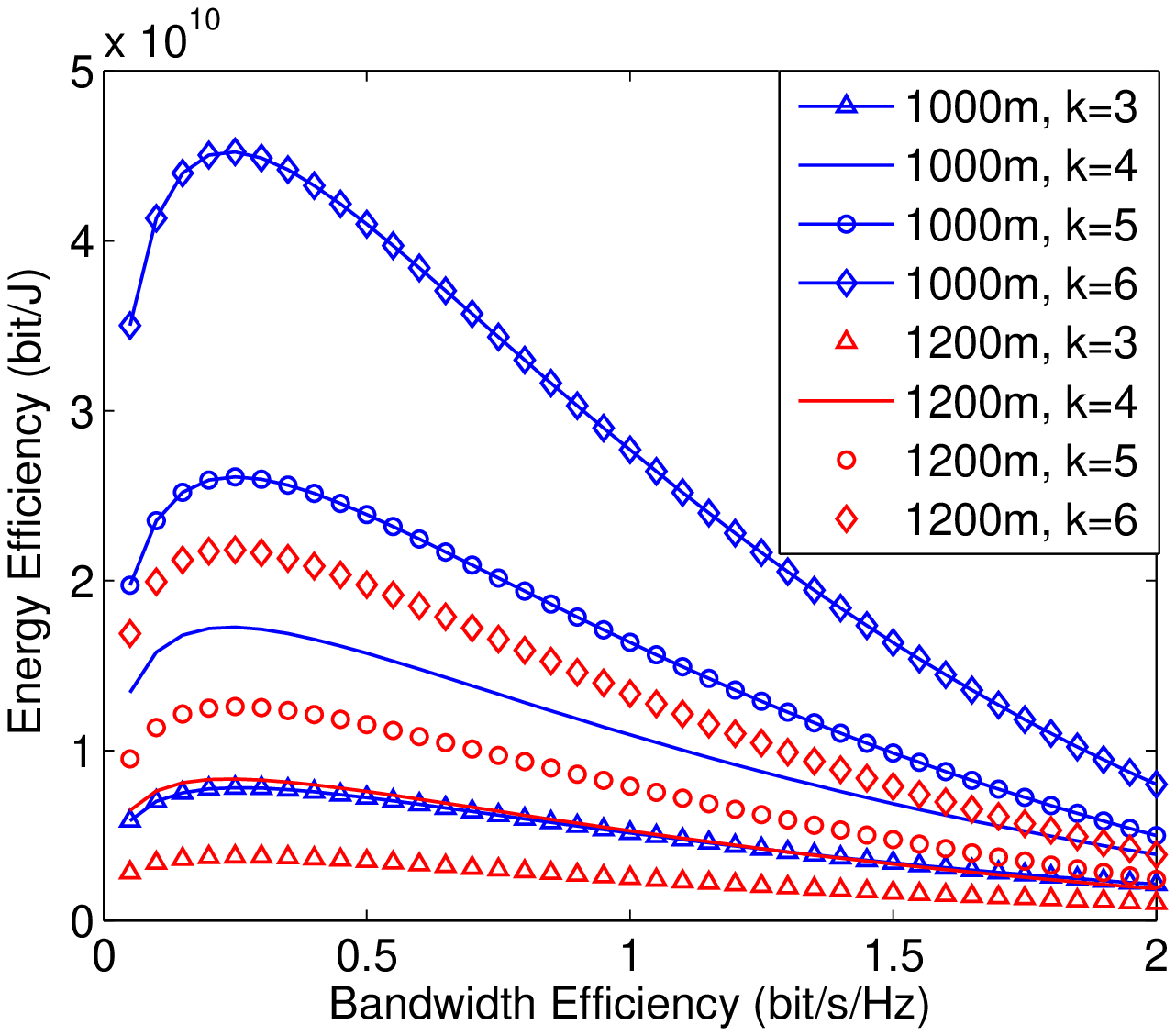}}
\caption{EE-BE relation for different number of relays when $\alpha=4$, processing energy is ignored, (a) $d_{route}= 1000$ m, $k$ varies from $0$ to $6$ (b) $d_{route}= 1000$ m or $1200$ m, $k$ varies from $3$ to $6$.}
\end{figure}
\indent
In Fig.~\ref{txpower_k06}, we plot EE and BE along a route with length $1000$ m and $k$ varies from $0$ to $6$. Again, we ignore the processing energy here, so only transmission energy is considered in this figure. Given the length of a transmission path and a certain BE, EE will increase with increased number of equidistant relays. In other words, if two routes connecting A and B have the same length, then the route with more relays consumes less transmission energy while still achieves the desired end-to-end rate (i.e., BE). This can be explained as follows. Given the length of a path, more relays along that path means smaller distance in each hop, and hence smaller transmission energy consumption in each hop. This effect dominates the increase in energy consumption resulting from more relays are consuming energy.\\
\indent
Fig.~\ref{txpower_compa_k36} illustrates how length of the route affects its performance. Given BE and the total number of relays, EE decreases as the length of the route increases, since longer distance needs larger transmission power to maintain the same transmission rate.
Combining Fig.~\ref{txpower_compa_k36} and Fig.~\ref{txpower_k06}, we conclude a general principle when comparing different routes in terms of EE and BE: given end-to-end transmission rate, route with shorter length and more relays tends to consume less energy. This principle has a key assumption that processing energy is small compared with transmission energy. But this is not always true: processing energy may dominate the total energy consumption at low transmission rate.\\
\indent
Fig.~\ref{total_large_proc} and Fig.~\ref{total_small_proc} depict EE and BE relation for $P_{0,0}=5\times 10^{-6}$ and $P_{0,0}=5\times 10^{-7}$ mJ/channel use, respectively. With a high processing energy, e.g., in Fig.~\ref{total_large_proc}, given the length of the route, a route with two relays in between has the best EE for most transmission rates. The main reason is that $P_{proc,2}$ is smallest among different $P_{proc,k}$s, $k=1,...,6$. If processing energy is small, e.g., in Fig.~\ref{total_small_proc}, then the EE-BE relation resembles that in Fig.~\ref{txpower_k06}, in the sense that a route with more relays tend to has larger EE as BE gets higher. Besides, if the number of the relays are given, the transmission rate which gives the maximal EE is different for different $k$. When $k=2$, the system can achieve best EE by transmitting at an end-to-end rate around $1.1$ bit/channel use.\\

\begin{figure}
\centering
\includegraphics[scale=0.6]{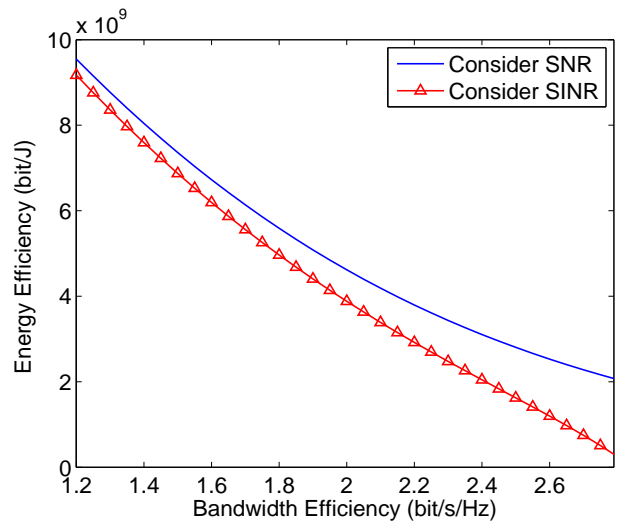}
\caption{Comparison of EE-BE relation with (SINR) and without (SNR) considering interference: $d_{route} = 1000$ m, $\alpha=4$, $k=4$, processing energy is ignored
.}\label{comparison_SNR_SINR}
\end{figure}
\indent
Fig.~\ref{comparison_SNR_SINR} explains the reason why we cannot ignore the interference influence when using capacity formula to derive EE and BE relation. The error percentage when using SNR instead of SINR to compute capacity increases from  4.1\% at BE$=1.2$ bit/s/Hz to 325\% at BE$=2.75$ bit/s/Hz. Except the large difference in the numerical results of EE, there is another drawback when we only consider SNR, that is, it cannot present the upper limit of the achievable transmission rate. If we use SNR in the capacity formula, we can achieve any rate as long as the transmission power is large enough, however, in reality it is not true. Taking $k=4$ as an example, in (\ref{five_1}), $P_{R_2}$, $P_{R_4}$ and $P_{B}$ are of the same order, so there is an upper limit of $R$ on the left-hand side.\\
\begin{figure}
\centering
\includegraphics[scale=0.6]{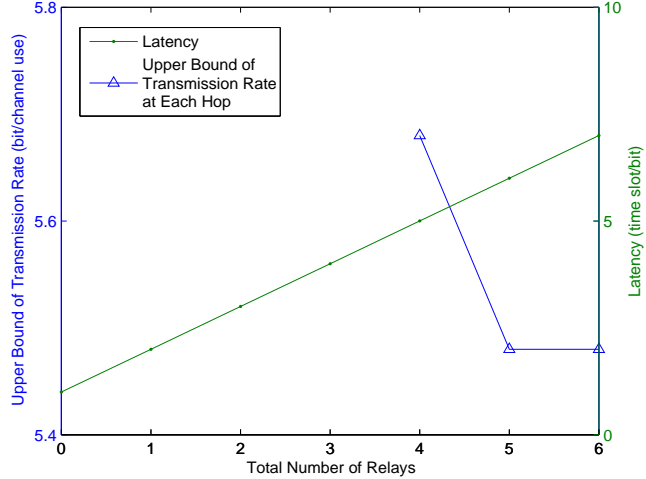}
\caption{Upper limit of the transmission rate and latency for various number of relays where $d_{route} = 1000$ m, $\alpha=4$.}\label{rate_limit_latency}
\end{figure}
\indent
In Fig.~\ref{rate_limit_latency} we plot the upper bound of the transmission rate as well as latency for different $k$s. The reason for bounded transmission rate is that the interference from other nodes is of the same order as the transmission power, resulting from the common rate transmission scheme. Since there is no interference in the transmission pattern when $k\le3$, no upper limit of rate exists. Previous analysis, i.e., Fig.~\ref{txpower_k06}, shows that more relays along a path can decrease the total transmission energy, but there is one big disadvantage associated with more relays, that is, latency. For a route with k relays, latency is $k+1$ time slot/bit, since each bit will be forwarded by $k+1$ hops before it reaches destination. It is different from the time which the two end nodes spend in receiving consecutive packets from each other, which is 2 time slots for $k\le1$ and 4 time slots for $k\ge2$.\\
\begin{figure}
\centering
\subfigure[]{\label{objective_large_proc}
\includegraphics[scale=0.5]{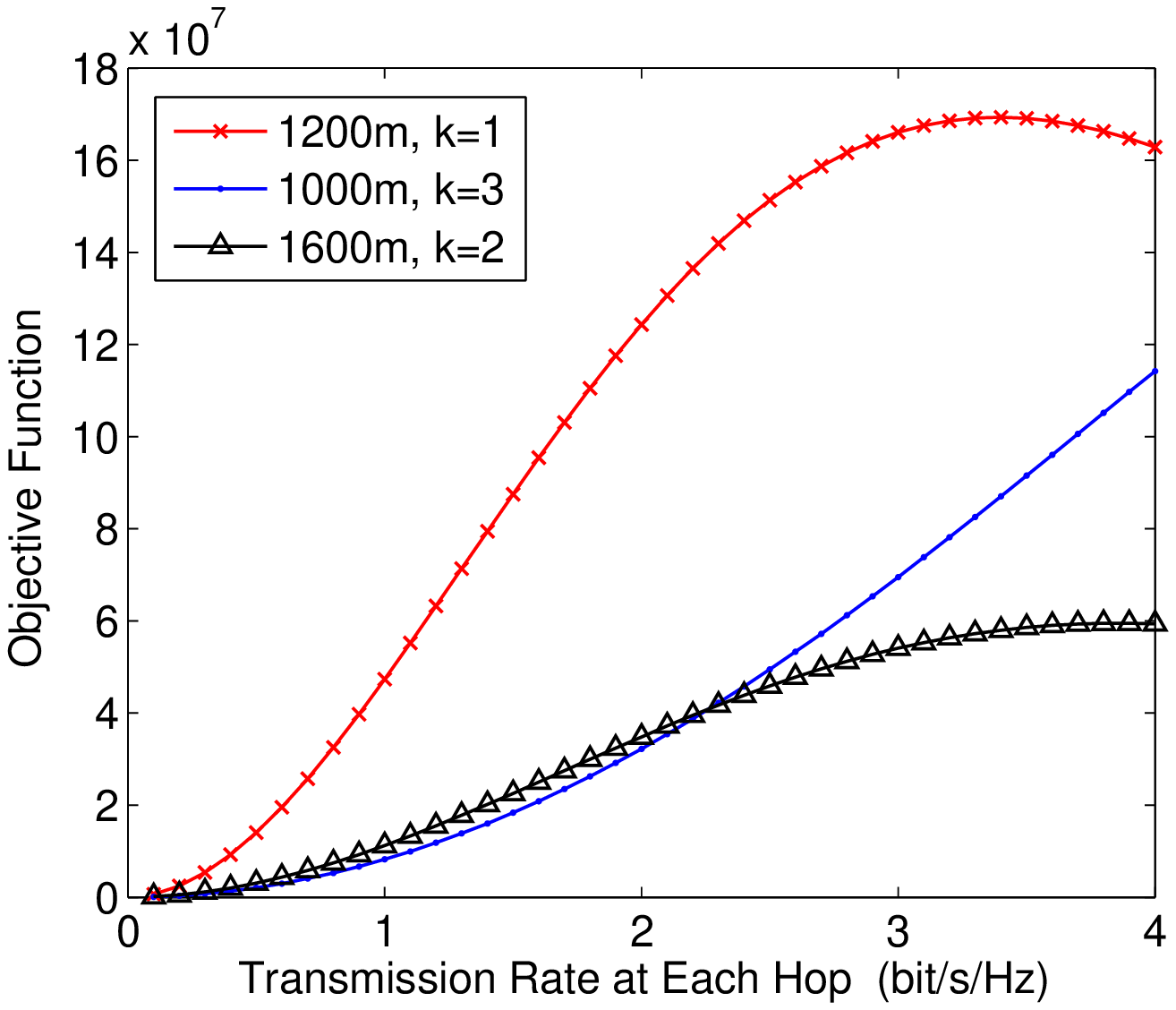}}
\subfigure[]{\label{objective_small_proc}
\includegraphics[scale=0.5]{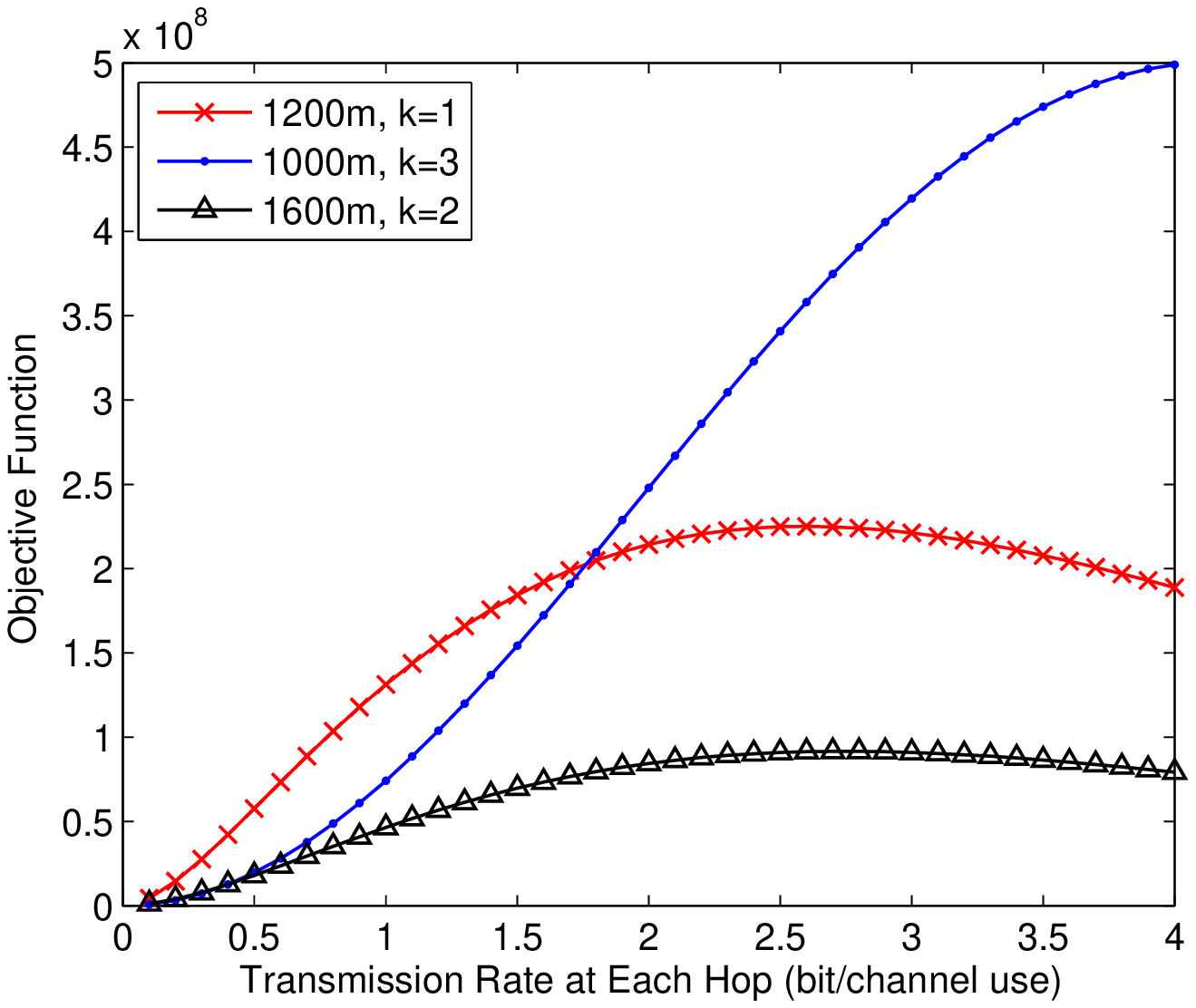}}
\caption{Comparison of the three routes in Fig.~\ref{model}: $\alpha=4$, $\eta=0.75$, (a) $P_{0,0}=5\times 10^{-6}$ (b) $P_{0,0}=5\times 10^{-7}$.}
\end{figure}
\indent
Previous analysis, i.e., Fig.~\ref{txpower_k06}, tells us that more relays of a path tends to decrease the total transmission energy. However, more relays incurs larger latency. Thus, we need to jointly consider the three performance metrics, in order to decide which one of the three routes in Fig.~\ref{model} performs best. A route with high EE, high BE and low latency is preferred, so a general performance metric can be defined as
\begin{equation}
F=\frac{\text{EE}/\text{EE}_{\max}\times \text{BE}/\text{BE}_{\max}}{\text{latency}/\text{latency}_{\max}}=\frac{\text{EE}\times \text{BE}}{\text{latency}}\times K, \label{metric}
\end{equation}
where $\text{EE}_{\max}$, $\text{BE}_{\max}$, and $\text{latency}_{\max}$ are the maximum achievable EE, BE, and latency of a system, e.g., Fig.~\ref{model}; and $K$ is a constant formed by the three maximum values. $F$ is an objective function that we want it to be as large as possible. Note that the definition of $F$ is not unique, e.g., if one cares more about EE, a different $F$ function can be designed so that route with a large EE tends to have a high $F$ value.\\
\indent
Fig.~\ref{objective_large_proc} and Fig.~\ref{objective_small_proc} depict $F/K$ for the three routes in Fig.~\ref{model}, assuming their configurations (length and number of relays) are $1200$ m with $k=1$, $1000$ m with $k=3$, and $1600$ m with $k=2$.
When processing energy plays a dominant role, e.g., in Fig.~\ref{objective_large_proc}, the first route has the best performance, since it has the lowest latency and small processing energy consumption. When processing energy is smaller compared with transmission energy, e.g., in Fig.~\ref{objective_small_proc}, the advantage of the second route becomes more obvious as the transmission rate increases. If processing energy can be ignored, according to the general principle we¡¯ve analyzed previously, the second route has the best BE-EE relation since it has both the shortest length and the largest number of relays.
\section{Insights On the Routing Path Selection}
\subsection{Summary of the Previous Results}
\begin{itemize}
\item Given BE, whether relay in the middle point achieves the highest EE is decided by (\ref{case}). When the path loss exponent is larger than the threshold, relay is best located in the middle, besides, TWRC is more energy efficient than direct transmission.
\item TWRC, unlike normal channels, displays a different EE and BE relation: they are positive related at low BE and then involved in a tradeoff relation. Therefore, when rate is small, we can both decrease energy and increase rate at the same time.
\item Route with more relays tend to consume less transmission energy.
\item A fundamental limit of the transmission rate exists when relays' number $k\ge4$, due to the symmetry of the transmission scheme and the interference from other nodes.
\end{itemize}
\subsection{Guidelines on the design of routing strategy}
\indent
A joint consideration of the transmission rate, path loss exponent, the length of a route and the number of relays is necessary, in order to select a path with the best performance in terms of EE, BE and latency. The transmission rate and path loss exponent are utilized to determine whether TWRC performs better than direct transmission, and whether relay in the middle gives the best performance. If the transmission rate is low, then the processing energy is dominant, a route with a small number of relays tends to offer the best performance. As rate increases, transmission energy becomes dominant, then a route with shorter length and more relays tends to consumes less energy at the cost of higher latency.
\section{Conclusion}
In this paper we presented an information theoretical framework to study routing path selection for amplify-and-forward two-way relay networks. We formulated bandwidth efficiency, energy efficiency and latency for routes with different number of relays, assuming a simple Hop-by-Hop scheduling scheme. We also determine the optimal power allocation scheme that allows a route to consume the minimal energy while still achieve the same end-to-end transmission rate. Our theoretical formulations and simulation results help provide guidelines towards routing protocol design. Future work of interest includes considering routing path selection under the assumptions of multiple frequencies and imperfect scheduling.


\begin{thebibliography}{1}
\bibitem{proceedings}
B. Nazer and M. Gastpar, "Reliable Physical Layer Network Coding," in \emph{Proc. IEEE}, vol. 99, no. 3, pp. 438-460, March 2011.
\bibitem{hottopic}
S. Zhang, S C Liew, and P. P. Lam, "Hot Topic: Physical-Layer Network Coding," in \emph{Proc. ACM MobiCom}, pp. 358-365, Sept. 2006.
\bibitem{full}
B. Rankov and A. Wittneben, "Achievable Rate Regions for the Two-way Relay Channel," in \emph{Proc. IEEE Int. Symp. Inf. Theory (ISIT)}, pp. 1668-1672, July 2006.
\bibitem{embracing}
S. Katti, S. Gollakota, and D. Katabi, "Embracing Wireless Interference: Analog Network Coding," in \emph{Proc. ACM SIGCOMM}, Aug. 2007.
\bibitem{half}
S. J. Kim, N. Devroye, P. Mitran, and V. Tarokh, "Achievable rate regions and performance comparison of half duplex bi-directional relaying protocols," \emph{IEEE Trans. Inf. Theory}, vol. 57, no. 10, pp. 6405-6418, Oct. 2011.
\bibitem{compute}
B. Nazer and M. Gastpar, "Compute-and-Forward: Harnessing Interference Through Structured Codes," \emph{IEEE Trans. Inf. Theory}, vol. 57, no. 10, pp. 6463-6486, Oct. 2011.
\bibitem{line}
Q. You, Z. Chen, and Y. Li, "A Multihop Transmission Scheme With Detect-and-Forward Protocol and Network Coding in Two-Way Relay Fading Channels," \emph{IEEE Trans. Veh. Technol.}, vol. 61, no. 1, pp. 433-438, Jan. 2012.
\bibitem{star}
J. H. S$\o$rensen, R. Krigslund, P. Popovski, T. K. Akino, and T. Larsen, "Scalable DeNoise-and-Forward in bidirectional relay networks," \emph{Computer Networks}, vol. 54, no. 10, pp. 1607¨C1614 July, 2010.
\bibitem{denoising_map}
T. K. Akino, P. Popovski, and V. Tarokh, "Denoising Maps and Constellations for Wireless Network Coding in Two¨CWay Relaying Systems," in \emph{Proc. IEEE Global Telecommun. Conf. (Globecom)}, Dec. 2008.
\bibitem{anti}
P. Popovski and H. Yomo, "The Anti-Packets Can Increase the Achievable Throughput of a Wireless Multi-Hop Network," in \emph{Proc. IEEE Int. Conf. Commun. (ICC)}, June 2006.
\bibitem{multihop}
P. Popovski and H. Yomo, "Bi-directional Amplification of Throughput in a Wireless Multi-hop Network," \emph{Proc. IEEE 63rd Vehicular Technology Conf. (VTC 2006-Spring)}, pp. 588-593, May 2006.
\bibitem{AF-ANC-Even}
Q. You, Z. Chen, Y. Li and B. Vucetic, "Multi-hop Bi-directional Relay Transmission Schemes Using Amplify-and-Forward and Analog Network Coding," in \emph{Proc. IEEE Int. Conf. Commun. (ICC)}, June 2011.
\bibitem{EXT}
D. De Couto, D. Aguayo, J. Bicket, and R. Morris, "A High-Throughput Path Metric for Multihop Wireless Routing," in \emph{Proc. ACM MobiCom}, pp. 134-146, 2003.
\bibitem{ETT}
R. Draves, J. Padhye, and B. Zill, "Routing in Multiradio, Multihop Wireless Mesh Networks," in \emph{Proc. ACM MobiCom}, pp. 114-128, 2004.
\bibitem{EE-BE}
C. Bae, and W. E. Stark, "End-to-End Energy-Bandwidth Tradeoff in Multihop Wireless Networks," \emph{IEEE Trans. Inf. Theory}, vol. 55, no. 9, pp. 4051-4066, Sept. 2009.
\bibitem{EE-BE-MIMO}
C. L. Chen, W. E. Stark, and S. G. Chen, "Energy-Bandwidth Efficiency Tradeoff in MIMO Multi-Hop Wireless Networks," \emph{IEEE J. Sel. Areas Commun.}, vol. 29, no. 8, Spet. 2011.
\bibitem{ismore}
C. Sun, and C. Yang, "Is Two-way Relay More Energy Efficient?" in \emph{Proc. IEEE Global Telecommun. Conf. (Globecom)}, Dec. 2011.
\bibitem{verdu}
S. Verd$\acute{u}$, "Spectral Efficiency in the Wideband Regime," \emph{IEEE Trans. Inf. Theory}, Vol. 48, No. 6, pp. 1319-1343, June 2002.
\bibitem{neely}
M. J. Neely, "Optimal Energy and Delay Tradeoffs for Multi-User Wireless Downlinks," \emph{IEEE Trans. Inf. Theory}, vol. 53, no. 9, pp. 1-17, Sept. 2007.
\bibitem{outage}
Z. Yi, M. Ju, and I. M. Kim, "Outage Probability and Optimum Power Allocation for Analog Network Coding," \emph{IEEE Trans. Wireless Commun.}, vol. 10, no. 2, pp. 407-412, Feb. 2011.
\bibitem{provision}
Y. Li, X. Zhang, M. Peng, and W. Wang, "Power Provisioning and Relay Positioning for Two-Way Relay Channel With Analog Network Coding," \emph{IEEE Singal Processing Letters}, vol. 18, no. 9, pp. 517-520, Sept. 2011.
\bibitem{processing}
Q. Wang, M. Hempstead, and W. Yang, "A Realistic Power Consumption Model for Wireless Sensor Network Devices," in \emph{Proc. IEEE SECON}, pp. 286-295, Sept. 2006.
\bibitem{802.11a}
IEEE 802.11, \emph{Wireless LAN Medium Access Control (MAC) and Physical Layer (PHY) Specifications}, Standard, IEEE, Aug. 1999.
\end{thebibliography}
\end{document}